\documentclass[times]{cpeauth}

\usepackage{moreverb}
\usepackage[colorlinks,bookmarksopen,bookmarksnumbered,citecolor=red,urlcolor=red]{hyperref}
\usepackage[utf8]{inputenc} 

\newcommand\BibTeX{{\rmfamily B\kern-.05em \textsc{i\kern-.025em b}\kern-.08em
T\kern-.1667em\lower.7ex\hbox{E}\kern-.125emX}}

\def\volumeyear{2012} 

\usepackage{amsmath}
\usepackage{amssymb}
\usepackage{latexsym}

\usepackage{subfigure}
\usepackage{color}                      
\usepackage{listings}

\usepackage{bm}
\usepackage{algorithmic}
\usepackage[ruled]{algorithm}

\usepackage{mathtools}
\usepackage{graphicx}

\begin{document}
\runningheads{M. Szydlarski, P. Esterie, J. Falcou, L. Grigori, R. Stompor}{Parallel Spherical Harmonic Transforms on heterogeneous architectures}

\author{{\normalsize Mikolaj Szydlarski\affil{1}\corrauth, Pierre Esterie\affil{2}, Joel Falcou\affil{2}, Laura Grigori\affil{3,4}\corrauth and Radek Stompor\affil{4}\corrauth }}

\address{\affilnum{1} INRIA Saclay-\^Ile de France, F-91893 Orsay, France 
\break \affilnum{2}Universit\'e Paris Sud, F-91405 Orsay, France 
\break \affilnum{3}INRIA Rocquencourt, Alpines, B.P. 105, F-78153, Le Chesnay Cedex, France
\break \affilnum{4}UPMC Univ Paris 06, CNRS UMR 7598, Laboratoire Jacques-Louis Lions, F-75005, Paris, France
\break \affilnum{5} APC, Univ Paris Diderot, CNRS/IN2P3, CEA/Irfu, Obs de Paris, Sorbonne Paris Cit\'e, France}

\corraddr{mikolaj.szydlarski@\{inria.fr,gmail.com\}, laura.grigori@inria.fr, radek@apc.univ-paris-diderot.fr}

\def\aalm{{$\bm{a}_{\ell m}$}}
\def\blm{{$\bm{\beta}_{\ell m}$}}
\def\miu{{$\bm{\mu}_{m}$}}

\def\s2hat{{\sc s$^2$hat}}
\def\alm2map{{\tt alm2map}}
\def\map2alm{{\tt map2alm}}

\def\M{{$\cal M$}}
\def\L{{$\cal L$}}

\def\deltam{{$\bm{\Delta}_{m}\ $}}

\def\apjl{ApJ}%
\def\apj{ApJ}%

\def\l#1{\left#1}
\def\r#1{\right#1}
\def\simlt{\lower.5ex\hbox{$\; \buildrel < \over \sim \;$}}
\def\simgt{\lower.5ex\hbox{$\; \buildrel > \over \sim \;$}}
	
\newenvironment{indenteddescription}%
	{\begin{list}{}{\setlength{\labelwidth}{0pt}
	\setlength{\itemindent}{-\leftmargin}
	\setlength{\listparindent}{\parindent}
	\renewcommand{\makelabel}{\descriptionlabel}}}%
{\end{list}}

\title{Parallel Spherical Harmonic Transforms \\ on heterogeneous architectures (GPUs/multi-core CPUs)}

\begin{abstract}
	Spherical Harmonic Transforms (SHT) are at the heart of many scientific and practical applications ranging from  climate modelling to cosmological observations. In many of these areas new, cutting-edge science goals have been recently proposed requiring simulations and analyses of experimental or observational data at very high resolutions and of unprecedented volumes. Both these aspects pose formid\-able challenge for the currently existing implementations of the transforms.
	
This paper describes parallel algorithms for computing SHT with two
variants of intra-node parallelism appropriate for novel supercomputer
architectures, multi-core processors and Graphic Processing Units
(GPU). It also discusses their performance, alone and embedded within
a top-level, MPI-based parallelisation layer ported from the S$^2$HAT
library, in terms of their accuracy, overall efficiency and
scalability. We show that our inverse SHT run on GeForce 400 Series
GPUs equipped with latest CUDA architecture ("Fermi") outperforms the
state of the art implementation for a multi-core processor executed on
a current Intel Core i7-2600K.  Furthermore, we show that an MPI/CUDA
version of the inverse transform run on a cluster of 128 Nvidia Tesla
S1070 is as much as 3 times faster than the hybrid MPI/OpenMP version
executed on the same number of quad-core processors Intel Nehalem for
problem sizes motivated by our target applications.  Performance of
the direct transforms is however found to be at the best comparable in
these cases.  We discuss in detail the algorithmic solutions devised
for the major steps involved in the transforms calculation,
emphasising those with a major impact on their overall performance,
and elucidates the sources of the dichotomy between the direct and the
inverse operations.\\
\end{abstract}
\maketitle 


\section{Introduction}\label{sec:introduction} 

Spherical harmonic functions constitute an orthonormal complete basis for
signals defined on a 2-dimensional sphere. Spherical harmonic
transforms are therefore common in all scientific applications where
such signals are encountered. These include a number of diverse areas
ranging from weather forecasts and climate modelling, through
geophysics and planetology, to various applications in astrophysics
and cosmology. In these contexts a direct Spherical Harmonic Transform
(SHT) is used to calculate harmonic domain representations of the
signals, which often possess simpler properties and are therefore more
amenable to further investigation. An inverse SHT is then used to
synthesize a sky image given its harmonic representation. Both of
those are also used as means of facilitating the multiplication of
huge matrices, defined on the sphere, and having the property of being
diagonal in the harmonic domain. Such matrices play an important role
in statistical considerations of signals, which are statistically
isotropic and such operations are among key operations involved in Monte Carlo Markov
Chain sampling approaches used to study such signals,
e.g.~\cite{Tanner1992}.

\begin{figure}[htp]
  \begin{center}
    \includegraphics[width=0.6\textwidth]{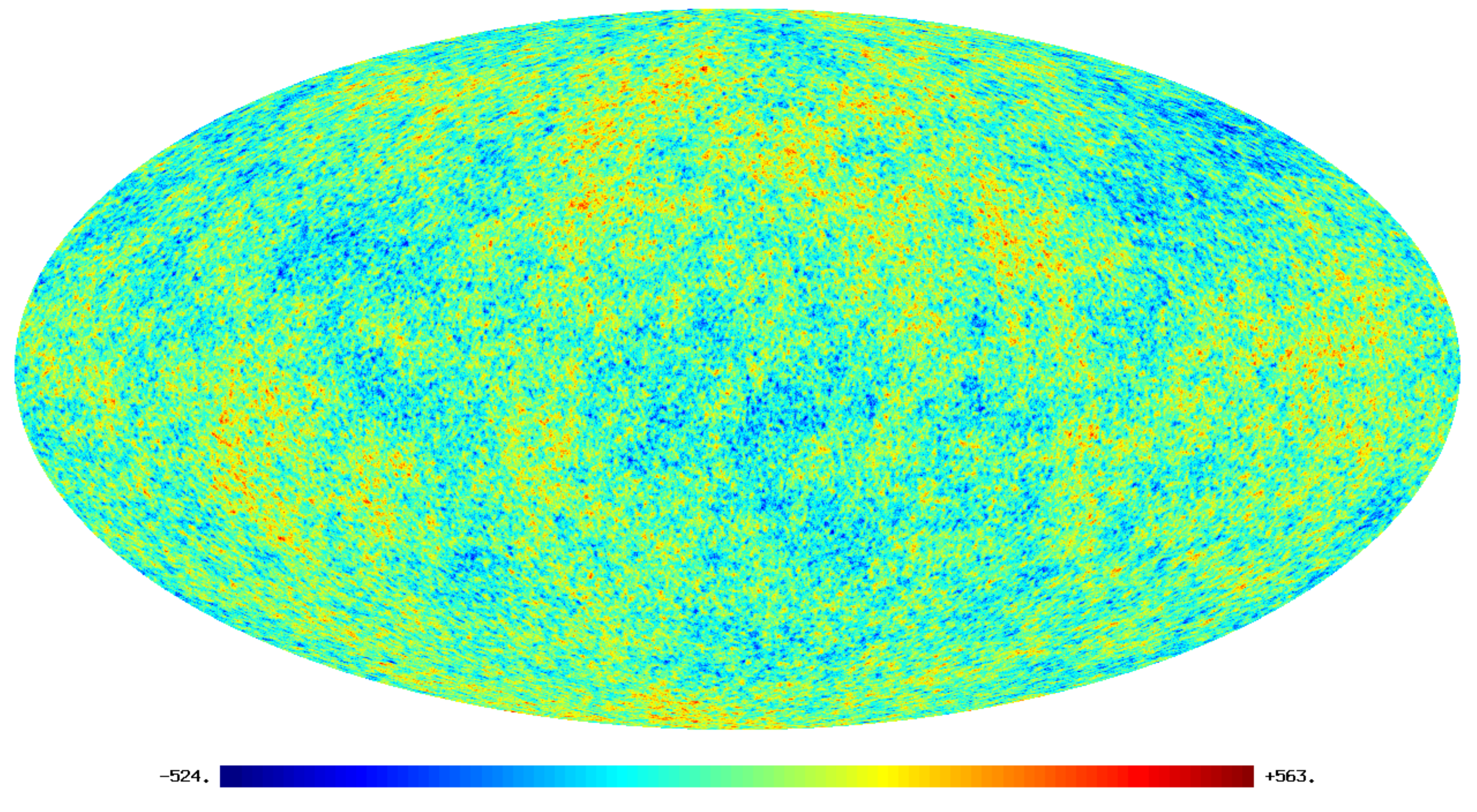}
  \end{center}
  \caption{Sky map example synthesised using our MPI/CUDA implementation of the \alm2map routine. The units are $\mu$K.
  An overall monopole term, with an amplitude of $\sim 2.7\,10^6\mu$K, has been subtracted to uncover the minute fluctuations.}\label{fig:skymap}
\end{figure}

The specific goal of this work is to assist simulation and analysis
efforts related to the Cosmic Microwave Background (CMB) studies. CMB
is an electromagnetic radiation left over after the hot and dense
initial phase of the evolution of the Universe, which is popularly
referred to as the Big Bang. Its observations are one of the primary
ways to study the Universe.  The CMB photons reaching us from afar
carry an image of the Universe at the time when it had just a small
fraction ($\sim 10^{-4}$) of its current age of $\sim 13$ Gyears.
This image confirms that the early Universe was nearly homogeneous and
only small, 1 part in $10^5$, deviations were present, as displayed in
Fig~\ref{fig:skymap}. Those initiated the process of so-called
structure formation, which eventually led to the Universe as we
observe today. The studies of the statistical properties of these
small deviations are one of the major pillars on which the present day
cosmology is built. On the observational front the CMB anisotropies
are targeted by an entire slew of operating and forthcoming
experiments. These include a currently observing European satellite
called Planck\footnote{{\scriptsize {\sc planck:}
    \url{http://sci.esa.int/science-e/www/area/index.cfm?fareaid=17}}},
which follows the footsteps of two very successful American missions,
COBE\footnote{{\scriptsize {\sc cobe:}
    \url{http://lambda.gsfc.nasa.gov/product/cobe/}}} and
WMAP\footnote{{\scriptsize {\sc wmap:}
    \url{http://map.gsfc.nasa.gov/}}}, as well as a number of
balloon-borne and ground-based observatories.  Owing to quickly
evolving detector technology, the volumes of the data collected by
these experiments have been increasing at the Moore's law rate,
reaching at present values on the order of petabytes. These new data
aiming at progressively more challenging science goals not only
provide, or are expected to do so, images of the sky with an
unprecedented resolution, but their scientific exploitation will
require intensive, high precision, and voluminous simulations, which
in turn will require highly efficient novel approaches to the
calculation of SHT.  For instance, the current and forthcoming
balloon-borne and ground-based experiments will produce maps of the
sky containing as many as ${\cal O}(10^5-10^6)$ pixels and up to
${\cal O}(10^6 -10^7)$ harmonic modes. Maps from already operating
Planck satellite will consist of between ${\cal O}(10^6)$ and ${\cal
  O}(10^8)$ pixels and harmonic modes. The production of high
precision simulations reproducing reliably polarized properties of the
CMB signal, in particular generated due to the so called gravitational
lensing, requires an effective number of pixels and harmonic modes, as
big as ${\cal O}(10^{9})$. These latter effects constitute some of the
most exciting research directions in the CMB area.  The SHT
implementation, which could successfully address all these needs would
not only have to scale well in the range of interest, but also be
sufficiently quick to be used as part of massive Monte Carlo
simulations or extensive sampling solvers, which are both important
steps of the scientific exploitation of the CMB data.

At this time there are a few software packages available which permit
calculations of the SHT. These include {\sc
  healpix}~\cite{healpix_www}, {\sc glesp}~\cite{glesp_www}, {\sc
  ccsht}~\cite{ccsht_www}, {\sc libpsht}~\cite{libpsht_www}, and {\sc
  s$^2$hat}~\cite{s2hat_www}, which are commonly used in the CMB
research, and some others such as, {\sc
  spharmonickit/s2kit}~\cite{spharmonickit_www} and {\sc
  spherpack}~\cite{spherpack_www}.  They propose different levels of
parallelism and optimization, while implementing, with the exception
of two last ones, essentially the same algorithm. Among those, the
{\sc libpsht} package~\cite{libpsht} offers typically the best
performance, due to efficient code optimizations, and it is considered
as the state of the art implementation on serial and shared-memory
platforms, at least for some of the popular sky pixelization schemes.
Conversely, the implementation offered by the {\sc s$^2$hat} library
is best adapted to distributed memory machines, and fully parallelized
and scalable with respect to both memory usage and calculation
time. Moreover, it provides a convenient and flexible framework
straightforwardly extensible to allow for multiple parallelization
levels based on hybrid programming models. Given the specific
applications considered in this work, the distributed memory
parallelism seems inevitable, and it is therefore the {\sc s$^2$hat}
package that we select as a starting point for this work.

The basic algorithm we use hereafter, and which is implemented in both
{\sc libpsht} and {\sc s$^2$hat} libraries and described in the next
section, scales as ${\cal O}({\cal R}_N\,\ell^2_{max})$, where ${\cal
  R}_N$ is the number of iso-latitudinal rings of pixels in the
analyzed map and $\ell_{max}$ fixes the resolution and thus determines
the number of modes in the harmonic domain, equal to $\simeq
\ell_{max}^2$. For full sky maps we have usually $\ell_{max}
\propto{\cal R}_N \propto n_{pix}^{1/2}$, where $n_{pix}$ is the total
number of sky pixels, and therefore the typical complexity is ${\cal
  O}(\ell_{max}^{3}) \propto {\cal O}(n_{pix}^{3/2})$.  We note that
further improvements of this overall scaling are possible.  For
instance, if two identical transforms have to, or can, be done
simultaneously, we could store precomputed data and speed up the
computation by a factor of up to $2$. However, in practice such
speedups are difficult to obtain due to very expensive operations of
reading or moving the precomputed data between different hierarchies
of memory.  In this paper we focus only on the core algorithm and
leave an investigation of such potential extensions as future work.

We note that alternative algorithms have been also proposed, some of
which display superior complexity. In particular, Driscoll and
Healy~\cite{Driscoll_Healy_1994} proposed a divide-and-conquer
algorithm with a theoretical scaling of ${\cal O}(n_{pix}\,\ln^2
n_{pix})$. This approach is limited to special equidistant sphere
grids and being inherently numerically unstable, it requires
corrective measures to ensure high precision results. This in turn
affects its overall performance. For instance, the software {\sc
  spharmonickit/s2kit}, which is the most widely used implementation
of this approach, has been found a factor 3 slower than the {\sc
  healpix} transforms implementing the standard ${\cal
  O}(n_{pix}^{3/2})$ method at the intermediate resolution of
$\ell_{max} \sim 1024$~\cite{Wiaux_etal_2006}. Other algorithms also
exist and typically involve a precomputation step of the same
complexity, but often less favourable prefactors, as the standard
approach, and an actual calculation of the transforms, which typically
exploits either the Fast Multipole Methods, e.g.,
~\cite{SudaTakami2002, Tygert2008} or matrix compression techniques,
e.g., ~\cite{Mohlenkamp1999, Tygert2010}, to bring down its overall
scaling to ${\cal O}(n_{pix}\ln n_{pix})$ ~\cite{SudaTakami2002},
${\cal O}(n_{pix}\ln^2 n_{pix})$~\cite{Tygert2008, Tygert2010}, or
${\cal O}(n_{pix}^{5/4} \ln n_{pix})$ ~\cite{Mohlenkamp1999}.  The
methods of this class typically require significant memory resources
needed to store the precomputation products and are advantageous only
if a sufficient number of successive transforms of the same type has
to be performed in order to compensate for the precomputation costs.
The most recent, and arguably satisfactory, implementation of such
ideas is a package called {\sc wavemoth}\footnote{{\scriptsize {\sc
      Wavemoth:
    }https://github.com/wavemoth/wavemoth}}~\cite{wavemoth}, which
achieves a speed-up of the inverse SHT with respect to the {\sc
  libpsht} library by a factor ranging from $3$ to $6$ for $\ell_{max}
\sim 4000$. In such a case the required extra memory is on the order
of $40$ GBytes and it depends strongly, i.e., $\propto \ell_{max}^3$,
on the resolution.  We also note that in some applications the need to
use SHT can be sometimes by-passed by resorting to approximate but
numerically efficient means such as the Fast Fourier Transforms as for
instance in the context of convolutions on the sphere as discussed in
~\cite{Elsner2011}.

In many practical applications, as the ones driving this research, SHT
are just one of many processing steps, which need to be performed to
accomplish a final task. In such a context, the memory consumption and
the ability of using large number of processors frequently emerge as
the two most relevant requirements, as the cost of the SHT transforms
is often subdominant and their complexity less important than that of
the other operations.  From the memory point of view, the standard
algorithm has the smallest memory footprint and therefore remains an
attractive option. The important question is then whether its
efficient, scalable, parallel implementation is indeed possible.  Such
a question is particularly pertinent in the context of heterogeneous
architectures and hybrid programming models and this is the context
investigated in this work.

The past work on the parallelisation of SHT includes a parallel
version of the two step algorithm of Driscoll and Healy introduced by
\cite{inda2001efficient}, the algorithm of \cite{Drake_algorithm888}
based on rephrasing the transforms as matrix operations and therefore
making them effectively vectorizable, and a shared memory
implementation available in the {\sc libpsht} package. More recently,
algorithms have been developed for
GPUs~\cite{gpu_soman,hupca2012}. This latter work provided a direct
motivation for the investigation presented here, which describes what,
to the best of our knowledge, is the first hybrid design of parallel
SHT algorithms suitable for a cluster of GPU-based architectures and
current high-performance multi-core processors, and involving hybrid
OpenMP/MPI and MPI/CUDA programming.  We find that once carefully
optimised and implemented, the algorithm displays nearly perfect
scaling in both cases, extending at least up to 128 MPI processes
mapped on the same number of pairs of multi-core CPU-GPU. We also find
that inverse SHT run on GeForce 400 Series GPGPUs equipped with the
latest CUDA device ("Fermi") outperforms a state of the art
implementation for multi-core processors executed on latest Intel Core
i7-2600K, while the direct transforms in both these cases perform
comparably.

This paper is organised as follows. In section \ref{sec:background} we
introduce the algebraic background of the spherical harmonic
transforms. In section~\ref{sec:basic_algorithm} we describe a basic,
sequential algorithm and list useful assumptions concerning the
pixelisation of a sphere, which facilitate a number of acceleration
techniques used in our approach. In the following section we introduce
a detailed description of our parallel algorithms along with two
variants suitable for clusters of GPUs and clusters of
multi-cores. Section \ref{sec:experiments} presents results for both
implementations and finally section \ref{sec:conclusions} concludes
the paper.



\section{Algebraic background}\label{sec:background} 

\subsection{Definitions and notations}

Spherical harmonic functions constitute a complete, orthogonal basis on the sphere. They can be therefore used to define a unique decomposition of any
 function defined on the sphere,  $f : S^{2}\rightarrow\mathbb{R}$, and depending
on two spherical coordinates: $\theta\in(0,\pi)$ (colatitude) and $\phi\in[0,2\pi)$ (longitude -- measured  counterclockwise about the positive $z$-axis from the positive $x$-axis),
{ \small
	\begin{eqnarray}\label{eq:direct_continous}
		f\l(\theta,\phi\r) & = & \sum_{\ell=0}^{\infty}\,\sum_{m=-\ell}^{\ell}\,\bm{a}_{\ell m}\,Y_{\ell m}\l(\theta,\phi\r),
	 \end{eqnarray}
}
where $\bm{a}_{\ell m}$ are spherical harmonic coefficients representing $f$ in the harmonic domain, $Y_{\ell m}$ is a spherical harmonic basis function of degree $\ell$ and order $m$. The spherical harmonic coefficients can be then computed by taking a scalar inner product of $f$ with the corresponding spherical harmonic, which can be expressed as a 2-dimensional integral
\begin{equation}\label{eq:invers_continous}
	\bm{a}_{lm}=\int\limits_{0}^{\pi}\,d\theta\, \int\limits_{0}^{2\pi}\,d\phi\;f(\theta,\phi)Y^{\dagger}_{lm}(\theta,\phi)\sin\theta,
\end{equation}
where $\dagger$ denotes complex conjugation.

In actual applications, the functions defined on the sphere are typically discretised and given only on a set of grid points. In this case,
the spherical harmonic transforms are used either to project the data, $\bm{s}_{n}$, onto the harmonic modes -- an operation referred to as an \emph{analysis} step or a direct transform -- or to reconstruct the discretised data given a set of harmonic coefficients, $\bm{a}_{\ell m}$ -- called an inverse transform or a {\em synthesis} step. These steps can be written as,

{ \small
	\begin{eqnarray}
		\tilde{\bm{a}}_{\ell m} & = & \sum_{\l\{\bm{\theta}_{n},\bm{\phi}_{n}\r\}}\,\bm{w}_{n}\,\bm{s}_{n}\l(\bm{\theta}_{n},\bm{\phi}_{n}\r)\,Y_{\ell m}\l(\bm{\theta}_{n},\bm{\phi}_{n}\r),\label{eqn:map2almDef} \\
		\tilde{\bm{s}}_{n}\l(\bm{\theta}_{n},\bm{\phi}_{n}\r) & = & \sum_{\ell=0}^{\ell_{max}}\,\sum_{m=-\ell}^{\ell}\,\bm{a}_{\ell m}\,Y_{\ell m}\l(\bm{\theta}_{n},\bm{\phi}_{n}\r).\label{eqn:alm2mapDef}
	 \end{eqnarray}
}
\noindent Given the choice of  a bandwidth, $\ell_{max}$, pixels weights, $\bm{w}_{n}$, and the grid geometry, both these transforms  may only be approximate, what is indicated by 
a tilde over the computed quantities. This is because of the sky sampling issues on the analysis step, or the assumed bandwidth in the case of the synthesis.

In CMB applications, $\bm{s}_{n}$ is typically a vector of
real-valued, pixelized data, e.g. the brightness of incoming CMB
photons, assigned to $n_{pix}$ locations, $\l(\theta_{n},
\phi_{n}\r)$, on the sky corresponding to centres of suitable chosen
sky pixels, and is referred to hereafter as {\em a map}. The parameter
$\ell_{max}$ defines the maximum order of the Legendre function and
thus a band-limit of the field $\bm{s}_{n}$, and in practice it is set
by the experimental resolution.

The basis functions $Y_{\ell m}\l(\bm{\theta}_{n},\bm{\phi}_{n}\r)$ are defined in terms of normalised associated Legendre functions, ${\cal P}_{\ell m}$, of degree $\ell$ and order $m$,
{ \small
	\begin{eqnarray}
		Y_{\ell m}\l(\bm{\theta}_{n},\bm{\phi}_{n}\r) \equiv {\cal P}_{\ell m}\l(\cos \bm{\theta}_{n}\r) e^{i m \bm{\phi}_{n}}. \label{eqn:YlmDef}
	\end{eqnarray}
}
As the associated Legendre functions  are real-valued, we have that  $Y_{\ell\, -m}=(-1)^{m}Y^{\dagger}_{\ell m}$, which we will use later on  to compute all spherical harmonics in terms of the associated Legendre functions $P_{\ell m}$ with $m \geq 0$. We note that from this we also have that $\bm{a}_{\ell m}^\dagger = (-1)^m\,\bm{a}_{\ell \, -m}$ for any real-valued data, $\bm{s}$.

Using equation~\eqref{eqn:YlmDef} we can separate the variables on the rhs of equations~(\ref{eqn:map2almDef}) and~(\ref{eqn:alm2mapDef})  thus reducing the computation of the spherical harmonic transform  to a  Fourier transform in the longitudinal coordinate $\phi_{n}$ followed by a projection onto the associated Legendre functions,
\begin{eqnarray}\label{eq:separation}
	\tilde{\bm{a}}_{\ell m} & = & \sum_{\l\{\bm{\theta}_{n}\r\}} \l[\l( \sum_{\l\{\bm{\phi}_{n}\r\}}\,\bm{w}_{n}\,\bm{s}_{n}\l(\bm{\theta}_{n},\bm{\phi}_{n}\r)\,e^{i m \bm{\phi}_{n}}\r){\cal P}_{\ell m}\l(\cos \bm{\theta}_{n}\r)\r].
\end{eqnarray}

The associated Legendre functions satisfy the following recurrence (with respect to the multipole number $\ell$ for a fixed value of $m$) critical for the algorithms developed in this paper,
	\begin{eqnarray}\label{eqn:assLegRec}
		{\cal P}_{\ell+2, m}\l(x\r)  = \beta_{\ell+2, m} x \,  {\cal P}_{\ell+1, m}\l(x\r) + 
		                             \frac{\beta_{\ell+2, m}}{\beta_{\ell+1, m}} {\cal P}_{\ell m}\l(x\r),
	\end{eqnarray}
where
	\begin{eqnarray}\label{eqn:betaDef}
		\beta_{\ell m} = \sqrt{ \frac{4 \,\ell^2-1}{\ell^2-m^2}}.
	\end{eqnarray}
The recurrence is initialised by the starting values,
	\begin{eqnarray}\label{eqn:pmm}
		{\cal P}_{m m}\l(x\r) & = & \frac{1}{2^m\,m!}\, \sqrt{\frac{\l(2m+1\r)!}{4\pi}}\,\l(1-x^2\r)^{m} \nonumber \\
		&\equiv& \mu_m \, \l(1-x^2\r)^{m}, \\ 
		{\cal P}_{m+1, m}\l(x\r) & = & \beta_{\ell+1,m}\,x\,{\cal P}_{mm}\l(x\r).\label{eqn:pm1m}
	\end{eqnarray}
The recurrence is numerically stable but a special care has to be
taken to avoid under- or overflow for large values of
$\ell_{max}$~\cite{Gorski_etal_2005} e.g., for increasing $m$ the
${\cal P}_{m m}$ values can become extremely small such that they can
no longer be represented by the binary floating-point numbers in IEEE
754\footnote{The IEEE has standardised the computer representation for
  binary floating-point numbers in IEEE 754. This standard is followed
  by almost all modern machines.} standard. However, since we have
freedom to rescale the values of ${\cal P}_{\ell m}$, we can use it to
avoid the problem.  Consequently, on each step of the recurrence, the
newly computed value of the associated Legendre function is tested and
rescaled together with the preceding value, if found to be too close
to the over- or underflow limits. The rescaling coefficients are kept
track of, e.g., accumulated in form of their logarithms, and used to
scale back all the computed values of ${\cal P}_{\ell m}$ at the end
of the calculation. This scheme is based on two facts. First, the
values of the associated Legendre functions calculated via the
recurrence change gradually and rather slowly on each step. Second,
their actual values as needed by the transforms are representable
within the range of the double precision values.

The specific implementation of these ideas used in all our algorithms
follows that of the \s2hat software and the {\cal HEALPix} package (at
least up to its version 2.0). It employs a precomputed vector of
values, sampling the dynamic range of the representable double
precision numbers and thus avoids any explicit computation of
numerically-expensive functions, such as logarithms and
exponentials. This scaling vector is used to compare the values of
${\cal P}_{\ell m}$ computed on each step of the recurrence, and then
it is used to rescale them if needed. For sake of simplicity, in this
paper we assume that this is an integral part of associated Legendre
function evaluation (independent of the platform architecture) and we
omit an explicit description of the algorithm used for rescaling. For
more details we refer to \cite{hupca2012} and the documentation of the
{\cal HEALPix} package~\cite{healpix_www}.

\subsection{Specialised formulation}\label{sec:additional_assumptions} 

For a general grid of $n_{pix}$ points on the sphere, a direct
calculation of any of the SHT has computational complexity ${\cal
  O}(\ell^{2}_{max} n_{pix})$, where the factor ${\cal O}(\ell_{max})$
reflects the cost of calculating all necessary associated Legendre
functions up to the order $\ell_{max}$ for each of $n_{pix}$ points on
the sphere. This is clearly prohibitive for problems in the range of
our interest. For instance, the current and forthcoming balloon-borne
and ground-based experiments will produce maps of sky containing as
many as $n_{pix}\in[10^5,10^6]$ pixels and up to ${\cal O}(10^6
-10^7)$ harmonic modes. Maps from the already operating Planck
satellite will consist of between ${\cal O}(10^6)$ and ${\cal
  O}(10^8)$ pixels and harmonic modes.  To improve on such complexity
we will make some additional geometrical constraints, which are to be
imposed on any permissible pixelization, and which ensure that the
numerical cost is down to ${\cal O}(n_{pix}^{1/2}\,\ell_{max}^2)$. We
note that nearly all spherical grids, which are currently used in CMB
analysis (for instance HEALPix and GLESP), as well as in many other
application areas, conform with these assumptions, which
are~\cite{Gorski_etal_2005},
\begin{itemize}
	\item the map pixels are arranged on ${\cal R}_N\approx\sqrt{n_{pix}}$ iso-latitudinal rings, where latitude of each ring is identified by a unique polar angle $\theta_{n}$ (for sake of simplicity hereafter we will refer to this angle by the ring index i.e.,  $r_{n} \rightarrow \theta_{n}$, where $n\in[0,{\cal R}_{N}]$)
	\item within each ring $r_{n}$, pixels must be equidistant ($\Delta \phi=${\sc const}), and therefore, $\phi_n = \phi_0 + n\,\Delta\,\phi$, though their number can vary from a ring to another ring.
\end{itemize}

The requirement of the iso-latitude distribution for all pixels helps to cut on the number of required floating point operations (FLOPs) as the associated Legendre function need now be calculated only once for each pixel ring. 

Taking into account these restrictions and the definition of the basis function, $Y_{\ell m}$ equation~\eqref{eqn:YlmDef},we can rewrite (adapted from \cite{Gorski_etal_2005}) the synthesis step from equation~\eqref{eqn:alm2mapDef} as
{ \small
	\begin{equation}\label{eqn:Delta2map}
		\bm{s}\l({r}_{n},\bm{\phi}_{n}\r)  \, =  \, \sum_{m=-\ell_{max}}^{\ell_{max}} \, e^{i m \phi_{0}}\,\bm{\Delta}^{A}_m\l({r}_{n}\r),
	\end{equation}
}
where $\bm{\phi}_{0}$ denotes the $\phi$ angle of the first pixel in the ring and $\bm{\Delta}^{A}_m\l({r}_{n}\r)$ denotes a set of functions defined as,
{ \small
	\begin{equation}\label{eqn:DeltaDef}
		\bm{\Delta}^{A}_m\l({r}_{n}\r) \equiv \l\{
			\begin{array}{l l}
				{\displaystyle \sum_{\ell=0}^{\ell_{max}}\,\bm{a}_{\ell 0} \, {\cal P}_{\ell 0}\l(\cos{r}_{n}\r),} & {\displaystyle m = 0,}\\
				{\displaystyle \sum_{\ell=m}^{\ell_{max}}\,\bm{a}_{\ell m}\,{\cal P}_{\ell m}\l(\cos{r}_{n}\r),} & {\displaystyle m > 0,}\\
				{\displaystyle \sum_{\ell=\l|m\r|}^{\ell_{max}}\,\bm{a}_{\ell \l|m\r|}^\dagger\,{\cal P}_{\ell \l|m\r|}\l(\cos{r}_{n}\r),} & {\displaystyle m < 0.}
			\end{array}
			\r.
	\end{equation}
}
And similarly for the analysis step, equation~\eqref{eqn:map2almDef},
{ \small
	\begin{equation}\label{eqn:almdef}
		\bm{a}_{\ell m} = \sum^{{\cal R}_{N}}_{n = 0} \, \bm{\Delta}^{S}_m\l({r}_{n}\r)\,{\cal P}_{\ell m}\l(\cos{r}_{n}\r), 
	\end{equation}
}
where
{ \small
	\begin{equation}\label{eqn:almfftdef}
		\bm{\Delta}^{S}_m\l({r}_{n}\r) \equiv \sum_{\l\{\phi_{n}\r\}}\, \bm{w}\,\bm{s}\l({r}_{n},\bm{\phi}_{n}\r)\,e^{-im\phi_{0}}.
	\end{equation}
}

From above, we can see that the spherical harmonic transforms can be
split into two successive computations: one calculating the associated
Legendre functions, equations~\eqref{eqn:DeltaDef}
and~\eqref{eqn:almdef}, and the other performing the series of Fast
Fourier Transforms (FFTs), equations~\eqref{eqn:Delta2map}
and~\eqref{eqn:almfftdef}~\cite{Muciaccia_etal_1997}. This splitting
immensely facilitates the implementation.  For instance, in order to
evaluate Fourier transforms we can use third-party libraries
addressing this problem. In our application, the number of samples (pixels) per ring may vary and its prime decomposition may involve large prime factors. A care has to be therefore taken in selecting an appropriate FFT library. For instance, the well-known, standard version of the {\tt fftpack} library \cite{Schw82} performs well only for data vectors of the length $N_{FFT}$, whose prime decomposition  contains the factors $2$, $3$, and $5$ and its numerical complexity degrades from ${\cal O}\l(N_{FFT} \log N_{FFT}\r)$ to ${\cal O}(N_{FFT}^{2})$ in other cases.  {\sc Healpix} and {\sc libpsht} use therefore its modified version implementing Bluestein's algorithm for FFTs with large prime factors, guaranteeing ${\cal O}\l(N_{FFT} \log N_{FFT}\r)$ complexity. In this work, we employ by default in all versions of our routines the {\tt FFTW} library \cite{frigo2005design}, which ensures similar
scaling, competitive performance and portability.

\subsection{Time consumption}\label{sub:t_consumption} 

\begin{figure}[htp]
  \begin{center}
		\includegraphics[width=0.75\textwidth]{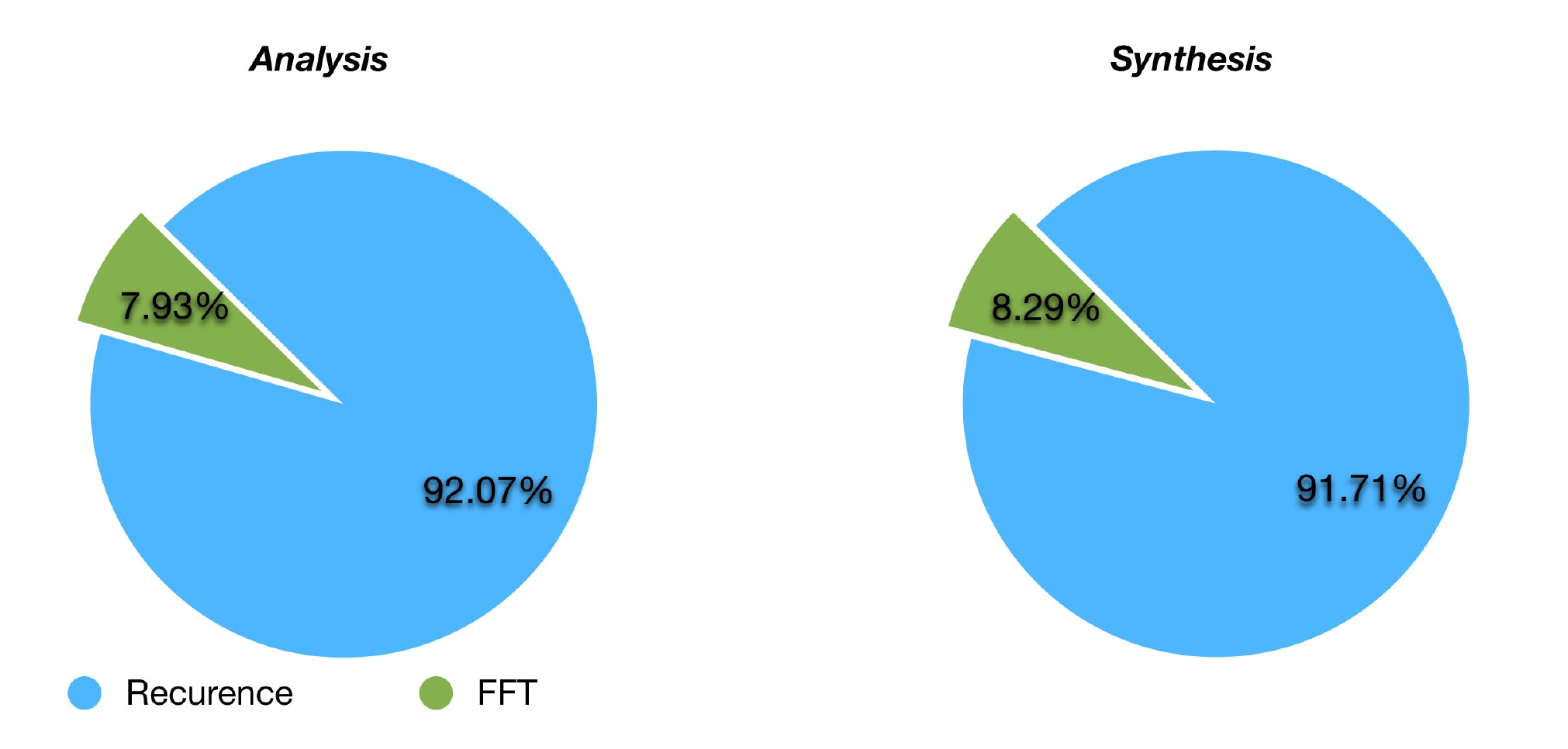}
  \end{center}
  \vskip -0.5truecm
  \caption{The overall time breakdown between two main steps of the SHT algorithm as computed assuming the same band limit ($\ell_{max} = 4096$) on a quad-core Intel i7-2600K processor.}\label{fig:pie_plot}
\end{figure}

Three parameters typically determine the data sizes and the complexity
of our problem. These are the total number of pixels, $n_{pix}$, the
number of rings of pixels in the analysed map, ${\cal R}_N$, and the
maximum order of the associated Legendre functions, $\ell_{max}$. In
particular the number of modes in the harmonic domain is then $\sim
\ell_{max}^2$.  In well defined cases these three parameters are not
completely independent. For example, in the case of full sky maps, we
have usually $\ell_{max} \propto{\cal R}_N \propto
n_{pix}^{1/2}$. This implies that the recurrence step requires ${\cal
  O}\l({\cal R}_{N}\ell_{max}^2\r)$ FLOPs, due to the fact that for
each of the ${\cal R}_{N}$ rings we have to calculate a full set, up
to $\ell_{max}$, of the associated Legendre functions, ${\cal P}_{\ell
  m}$, at cost $\propto \ell_{max}^2$. The Fourier transform step, as
mentioned above, has then complexity ${\cal O}\l({\cal
  R}_{N}\ell_{max} \log \ell_{max}\r)$, which should therefore be
subdominant as compared to the Legendre transforms. This is indeed
confirmed by our experimental results shown in
Figure~\ref{fig:pie_plot} depicting a typical breakdown of the average
overall time into the main steps of the SHT computation performed on
the quad-core Intel i7-2600K processor.
The evaluation of the Legendre transform takes nearly $10$ times longer than performing the FFTs. This motivated us to study the possible improvements of this part of the calculation. We explore this issue in the context of the heterogeneous architectures and hybrid programming models. In particular we introduce a parallel algorithm that is suitable for clusters of accelerators, namely multi-core processors and GPUs which we employ for an efficient evaluation of spherical harmonic transforms. 

\nocite{st_analysis}  


\section{Basic algorithm}\label{sec:basic_algorithm} 
In this section we present basic algorithms for the computation of the
spherical harmonic transforms, for sphere pixelizations which conform
with the assumptions listed in the previous section.  In
section~\ref{sec:parallelization} we then introduce their parallel
versions and discuss possible improvements and amendments as driven by
the specific architectures considered in this work.

Algorithm \ref{algo:map2almBasic} presents a computation of the
discrete, direct SHT, and thus implementing
equation~\eqref{eqn:map2almDef}, which for a given bandwidth,
$\ell_{max}$, and an input data vector, $\bm{s}$, computes a
corresponding set of the harmonic coefficients $\bm{a}_{\ell m}$.
\begin{algorithm}
\caption{{\sc Direct transform (equation~\eqref{eqn:map2almDef})}}
\begin{algorithmic}
{ \small \sl
	\REQUIRE{$\bm{s}\in\mathbb{R}_{n_{pix}}$ values}
	\STATE{ {\sc step 1 - $\bm{\Delta}^{S}_m\l(r\r)$ calculation} }
	\FOR{every ring $r$}
	\FOR{every $m = 0, ..., {m}_{max}$}
	\STATE{ \hspace{0.1cm} $\circ$ calculate $\bm{\Delta}^{S}_m\l(r\r)$ via FFT equation~\eqref{eqn:almfftdef}}
	\ENDFOR{($m$)}
	\ENDFOR{($r$)}
	\medskip
	\STATE { {\sc step 2 - Pre-computation }} \\
	\FOR{every $m = 0, ..., {m}_{max}$}
	\STATE { \hspace{0.1cm} $\circ$ $\bm{\mu}_m$ pre-computation equation~\eqref{eqn:pmm}\\ }
	\ENDFOR{ ($m$)}
	\medskip
   \STATE{ {\sc step 3 - Core calculation} } 
	\FOR{every ring $r$}
	\FOR{every $m = 0, ..., {m}_{max}$} 
	\FOR{every  $\ell = m, ..., \ell_{max}$} 
	\STATE{ \hspace{0.1cm} $\circ$ compute ${\cal P}_{\ell m}$ via equation~\eqref{eqn:assLegRec} }
	\STATE{ \hspace{0.1cm} $\circ$ update $\bm{a}_{\ell m}$, equation~\eqref{eqn:almdef} }
	\ENDFOR{ ($\ell$)}
	\ENDFOR{ ($m$)}
	\ENDFOR{ ($r$)}

   \medskip
	\RETURN {$\bm{a}_{\ell m}\in\mathbb{C}^{\ell\times m}$}
}
\end{algorithmic}
\label{algo:map2almBasic}
\end{algorithm}
Algorithm~\ref{algo:alm2mapBasic} describes a computation of the discrete, inverse transform as in equation~\eqref{eqn:alm2mapDef}, which takes as input a set of  harmonic coefficients, $\bm{a}_{\ell m}$, and returns a set of  real-valued numbers, $\bm{s}$.
\begin{algorithm}
\caption{{\sc Inverse Transform (equation~\eqref{eqn:alm2mapDef})}}
\begin{algorithmic}
{ \small \sl
	\REQUIRE {$\bm{a}_{\ell m}\in\mathbb{C}^{\ell\times m}$ values}
	\STATE { {\sc step 1 - Pre-computation }} \\
	\FOR{every $m = 1, ..., {m}_{max}$}
	\STATE { \hspace{0.1cm} $\circ$ $\bm{\mu}_m$ precomputation equation~\eqref{eqn:pmm}\\ }
	\ENDFOR{ ($m$)}
	\medskip
	\STATE { {\sc step 2 - $\bm{\Delta}^{A}_m$ calculation} }
	\FOR{every ring $r$}
	\FOR{every $m = 0, ..., {m}_{max}$} 
	\FOR{every  $\ell = m, ..., \ell_{max}$} 
	\STATE{ \hspace{0.1cm} $\circ$ compute ${\cal P}_{\ell m}$ via equation~\eqref{eqn:assLegRec} }
	\STATE{ \hspace{0.1cm} $\circ$ update $\bm{\Delta}^{A}_m\l( r\r)$, equation~\eqref{eqn:almfftdef} }
	\ENDFOR{ ($\ell$)}
	\ENDFOR{ ($m$)}
	\ENDFOR{ ($r$)}
	\medskip 
	\STATE{ {\sc step 3 - $\bm{s}$ calculation} }
	\FOR{every ring $r$}
	\FOR{every $m = 0, ..., {m}_{max}$}
	\STATE{ \hspace{0.1cm} $\circ$ calculate $\bm{s}$ via FFT and given  $\bm{\Delta}^{A}_m\l(r\r)$, equation~\eqref{eqn:Delta2map}}
	\ENDFOR{($m$)}
	\ENDFOR{ ($r$)}
   \medskip
	\RETURN {$\bm{s}\in\mathbb{R}_{n_{pix}}$}
}
\end{algorithmic}
\label{algo:alm2mapBasic}
\end{algorithm}
\subsection{Similarities}
The algorithms are almost the same in terms of the number and the type
of arithmetic operations involved in their execution (see for instance
Figure~\ref{fig:pie_plot} or a detailed analysis in
\cite{st_analysis}), employ a similar basic algorithmic principle and
have similar structure.  Indeed both these algorithms use a divide and
conquer approach, as they subdivide the initial problem, dealing with
the full set of pixels, into smaller subproblems of a similar form and
each involving only the rings of pixels.  Each of these subproblems
requires the calculation of the associated Legendre functions, which
is again solved by subdividing it into as many as $\ell_{max}$
recurrences with respect to the multipole number $\ell$ for a fixed
value of $m$ and only finally all the intermediate results are
combined together to provide the solution of the original problem.
Both these algorithms are also similar from the structural point of
view as they consist of three main steps:
\begin{itemize}
	\item pre-computation of the starting values of the recurrence \eqref{eqn:pmm} in the loop over all $m \in [0,m_{max}]$ ({\sc step} 2 in Algorithm \ref{algo:map2almBasic} and {\sc step} 1 in Algorithm \ref{algo:alm2mapBasic} ),
	\item evaluation of the associated Legendre functions for each ring of pixels via the recurrence from equation~\eqref{eqn:assLegRec}. This requires three nested loops, where the loop over $\ell$ is set as the innermost to facilitate the evaluation of the recurrence with respect to the multipole number $\ell$ for a fixed value of $m$ ({\sc step} 3 in Algorithm \ref{algo:map2almBasic} and {\sc step} 2 in Algorithm \ref{algo:alm2mapBasic} ).
	\item calculation of $\bm{\Delta}_m^{A/S}\l(\theta\r)$ using FFTs  ({\sc step} 1 in Algorithm \ref{algo:map2almBasic} and {\sc step} 3 in Algorithm \ref{algo:alm2mapBasic}).
\end{itemize}

\subsection{Differences}

The main difference is related to the dimension of the partial results
which need to be updated within the nested loop, where the associated
Legendre functions are evaluated ({\sc step} 3 in Algorithm
\ref{algo:map2almBasic} and {\sc step} 2 in Algorithm
\ref{algo:alm2mapBasic}). In the case of the direct transform, on
every pass of the loop over rings we need to update all the (non-zero)
elements of the matrix $\bm{a}_{\ell m}$ by the result of the product
between the object, $\bm{\Delta}^{S}_m$,
equation~\eqref{eqn:almfftdef}, which is precomputed earlier, and the
vector of the associated Legendre function with the azimuthal number
equal to $m$.  In the case of the inverse transform, the nested loop
({\sc step} 2 in Algorithm \ref{algo:alm2mapBasic}) is used to compute
$\bm{\Delta}^{A}_m$, equation~\eqref{eqn:DeltaDef}, that can be
expressed as the matrix-vector product,
\begin{equation}\label{eq:matrix_vec}
		\bm{\Delta}^{A}_m 
		= 
		\l(
			\begin{array}{c}
				\Delta^{A}_m(r_0,m)\\
				\Delta^{A}_m(r_1,m)\\
				\vdots\\
				\Delta^{A}_m({\cal R}_N,m)\\
			\end{array}
		\r) 
		=
		\l(
			\begin{array}{ccc}
				{\cal P}_{m m}\l(\cos r_{0}\r) & \hdots & {\cal P}_{\ell_{max} m}\l(\cos r_{0}\r) \\
				{\cal P}_{m m}\l(\cos r_{1}\r) & \hdots & {\cal P}_{\ell_max m}\l(\cos r_{1}\r) \\
				\vdots & \vdots & \vdots \\
				{\cal P}_{m m}\l(\cos r_{{\cal R}_{N}}\r) & \hdots & {\cal P}_{\ell_{max} m}\l(\cos r_{{\cal R}_{N}}\r) \\
			\end{array}
		\r)
		\,
		\l(
			\begin{array}{c}
				a_{m m} \\
				a_{m+1 m} \\
				\vdots\\
				a_{\ell_{max} m} \\
			\end{array}
		\r). 
\end{equation}

Depending on the specific computer architecture, this difference may
have a significant effect on the performance of these algorithms in
terms of runtime (see further investigation in this paper). We also
note that this asymmetry is due to using the recurrence with respect
to $\ell$ for the calculation of the associated Legendre functions,
which implies that the loop over $\ell$ should be placed as innermost,
what in turn prevents expressing the calculation of $\bm{a}_{\ell m}$
in Algorithm~\ref{algo:map2almBasic} as a matrix-vector operation.
 
\section{Parallel Spherical Harmonic Transforms}\label{sec:parallelization}

\subsection{Top-level distributed-memory parallel framework}

The need for a distributed memory parallelisation layer is driven by
the volumes of the current and anticipated data, both the maps and the
harmonic coefficients, and which have to be analysed in the
applications targeted here.  It is also driven by the required
flexibility of the transform implementations.  These are envisaged to
be frequently needed in massively parallel applications, where SHT
typically constitute a single step of an entire processing chain. This
requirement calls not only for a flexible interface and an ability to
use efficiently as many MPI processes as available, but also for a low
memory footprint per process.

The top level parallelism employed in this work is adopted from the
S$^2$HAT library originally developed by the last author. Though not
new, it has not been described in detail in the literature yet and we
present below its first detailed description including its performance
model and analysis. The framework is based on a distributed-memory
programming model and involves two computational stages separated by a
single instance of a global communication involving all the processes
performing the calculations. These together with a specialised, but
flexible data distribution, are the major hallmarks of the
framework. The two stages of the computation can be further
parallelised and/or optimised, and we discuss two examples of
potential intra-node parallelisation approaches below. We note that
the framework as such does not imply what algorithm should be used by
the MPI processes, e.g., ~\cite{wavemoth}, though of course its
overall performance and scalability will depend on the specific
choice.

\subsubsection{Data distribution and algorithm}

\begin{figure}[ht]
\centering
\subfigure[The iso-latitudinal rings distribution.]{
\includegraphics[width=0.55\textwidth]{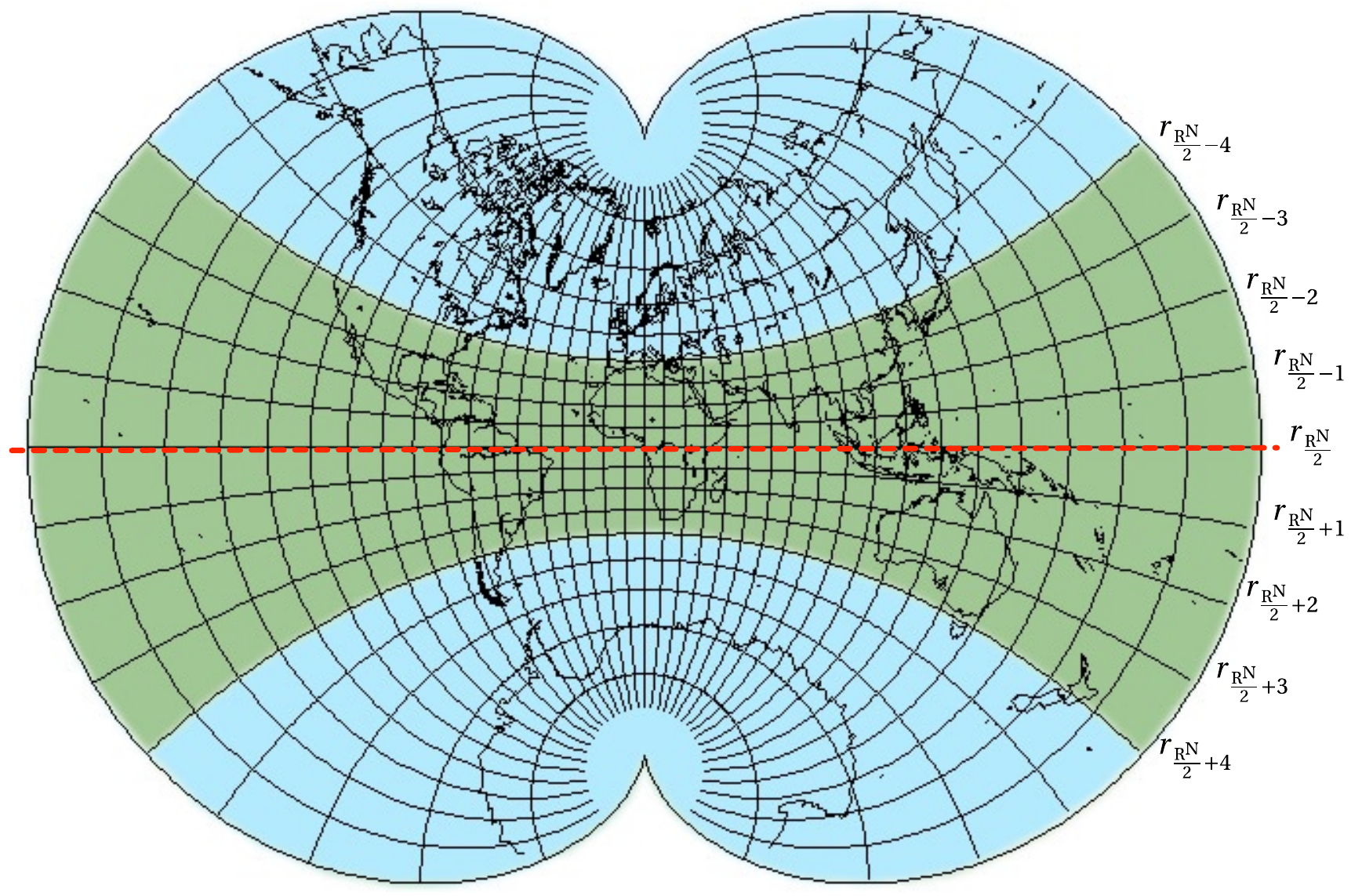}
\label{fig:dist_rings}
}
\subfigure[Distribution of ${a}_{\ell m}$ coefficients.]{
\includegraphics[width=0.37\textwidth]{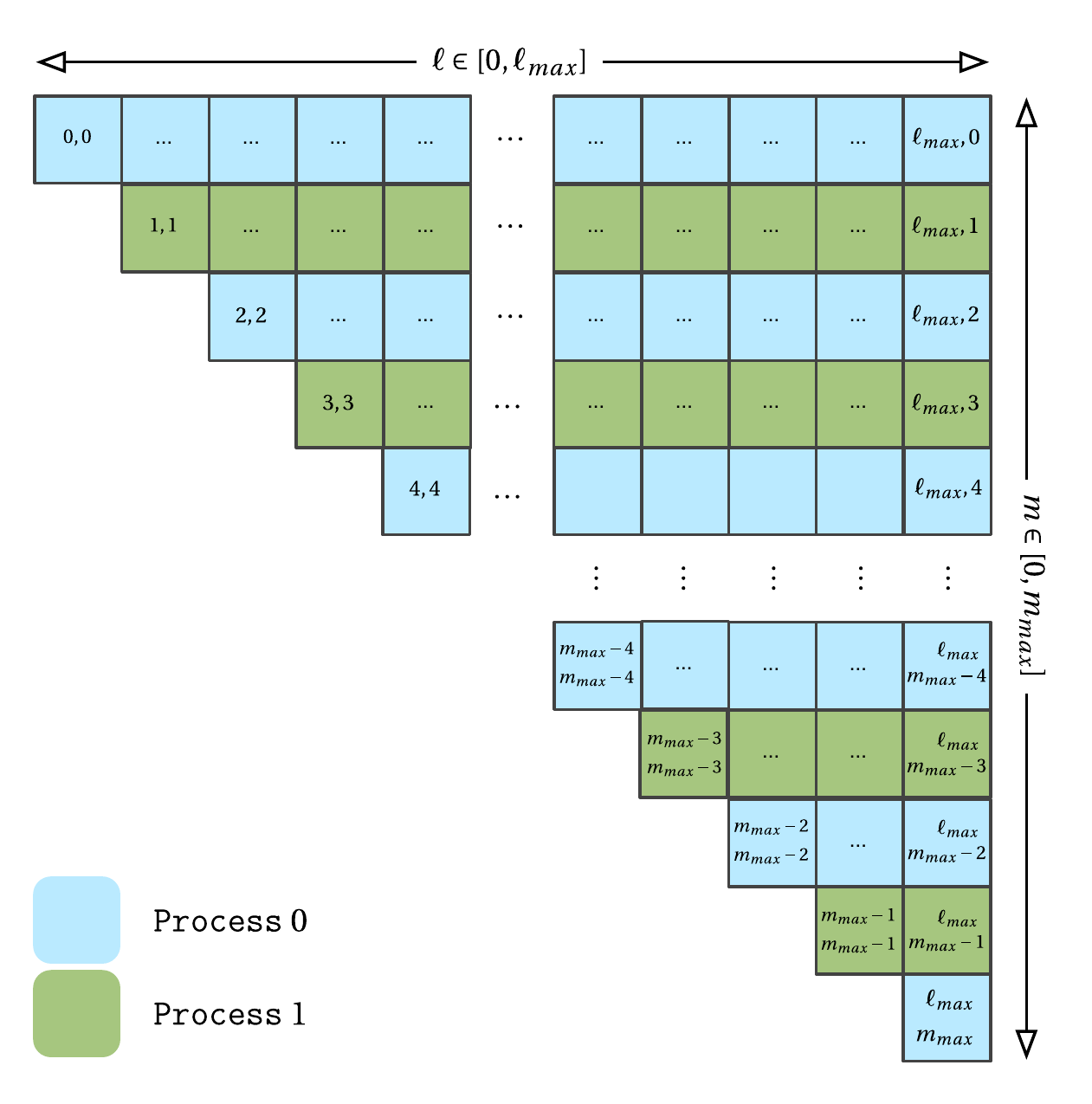}
\label{fig:dist_alm}
}
\caption{Example of the distribution of the iso-latitudinal rings and the  coefficients  $\bm{a}_{\ell m}$ among two processes. Colours mark elements assigned to a single process. For rings visualisation we used an August's Conformal Projection of the sphere on a two-cusped epicycloid.}
\end{figure}

The simple structure of the top-level parallelism outlined above is
facilitated by a data layout which assumes that both the map $\bm{s}$
and its harmonic coefficients $\bm{a}_{\ell m}$ are distributed. The
map distribution is ring-based, that is we assign to each process a
part of the full map consisting of the complete rings of pixels. To do
so we make use of the assumed equatorial symmetry of the pixelization
schemes and first divide all rings of one of the hemispheres
(including the equatorial ring if present) into disjoint subsets of
consecutive rings, and then each process is assigned one of the ring
subsets together with its symmetric counterpart from the other
hemisphere. If we denote the rings mapped on a processor $i$ as ${\cal
  R}_{i}$, then $\cup^{n_{procs}}_{i=0} {\cal R}_{i} = {\cal R}$,
where ${\cal R}$ is the set of all the rings and $n_{procs}$ is the
number of processes used in the computation. We also have ${\cal
  R}_{i}\cap{\cal R}_{j}=\emptyset$ if $i\ne j$. An example ring
distribution is depicted in Figure~\ref{fig:dist_rings}, where
different colours mark pixels on iso-latitudinal rings mapped on two
different processes.

In order to distribute the harmonic domain object we divide the 2-dimensional array of the coefficients $\bm{a}_{\ell m}$ with respect to a number, $m$, and assign to each process 
a subset of values of $m$, ${\cal M}_i$, and a corresponding subset of the hamonic coefficients, $\l\{ {a}_{\ell m} : m\in {\cal M}_i\  \hbox{and}\  \ell \in [m, \ell_{max}]\r\}$.  As with the ring subsets, the subsets of $m$ values have to be disjoint, ${\cal M}_i \cap {\cal M}_j = \emptyset$ if $i\ne j$, and all together include all the values, $\cup^{n_{procs}}_{i=0} {\cal M}_{i}={\cal M}$.
An example of the harmonic domain distribution is depicted in Figure~\ref{fig:dist_alm}.

This data layout allows to avoid any redundancy in terms of
computation and memory, and any need for an intermediate inter-process
communication.  At the same time it is general enough that at least in
principle can accommodate an entire slew of possibilities as driven by
real applications.  Any freedom in selecting a specific layout should
be used to ensure a proper load balancing in terms of both memory and
work, and it may, and in general will, depend on what specific
algorithm is used by each of the MPI processes. Though perfect load
balancing, memory- and work-wise, may not be always possible, in
practice we have found that good compromises usually exist. In
particular, the map distribution which assigns the same number of
rings to each process is typically satisfactory, even if it may in
principle lead to an unbalanced work and/or memory distribution. The
latter is indeed the case for the {\sc healpix} pixelization, which
does not preserve the same number of pixels per ring, what introduces
differences in the memory usage between the processes but also in the
workload. Some fine tuning could be used here to improve on both these
aspects, as it is indeed done in {\sc s$^2$hat}, but even without that
the differences are not critical. For the distribution of the harmonic
objects, each process $i$ stores a set of $m$ values defined as,
${\cal M}_i \equiv \l\{ m : m = i + k \, n_{procs} \ \hbox{or} \ m =
m_{max}-i-k\, n_{procs}, \ \hbox{where}\ k \in [0, m_{max}/2]\r\}$,
i.e., values of $m$ equal to $\l[i, i+n_{procs}, ...,
  m_{max}-i-n_{procs}, m_{max}-i\r]$.  This distribution appears
satisfactory in a number of cases as it is explained in the next
section.

We note that this data layout imposes an upper limit on the number of
processes which could be used, given by $\min ( m_{max}/2, {\cal
  R}_N/2)$ (we assign at least two rings and two $m$ values to each
process).  This is typically sufficient, as usually we have ${\cal
  R}_N \propto m_{max}\simeq \ell_{max}$, and the limit on the number
of processes is ${\cal O}\l(\ell_{max}\r)$.

With this data distribution, Algorithms~\ref{algo:map2almBasic}
and~\ref{algo:alm2mapBasic} require only straightforward modifications
to become parallel codes. As an example,
Algorithm~\ref{algo:Generallalm2mapS2HATParallel} details the parallel
inverse transform, which is a parallel version of
Algorithm~\ref{algo:alm2mapBasic}.
\begin{algorithm*} 
	\caption{{\sc General parallel \alm2map algorithm ( Code executed by each MPI process )}}
	\begin{algorithmic}
		\REQUIRE {$\bm{a}_{\ell m}\in\mathbb{C}^{\ell\times m}$ for $m\in{\cal M}_{i}$ and all $\ell$}
		\REQUIRE {indices of rings $r \in {\cal R}_i$}
		\medskip
		\STATE { {\sc step 1 - Pre-computation }} \\
		\FOR{every $m\in{\cal M}_{i}$}
		\STATE { \hspace{0.1cm} $\circ$ $\bm{\mu}_m$ pre-computation equation~\eqref{eqn:pmm}\\ }
		\ENDFOR{ ($m$)}
		\medskip
		
		\STATE { {\sc step 2 - $\bm{\Delta}^{A}_m$ calculation }} \\		
                  \FOR{ \underline{every} $m \in {\cal M}_i$}		
                      \STATE{
                     	$\circ$ precompute recurrence coefficients, $\beta_{\ell m}$~equation~\eqref{eqn:betaDef} \\
                      }
	   	    \FOR{ \underline{every} ring $r\in{\cal R}$}
			\STATE {
				$\circ$ initialise the recurrence: ${\cal P}_{mm}$, ${\cal P}_{m+1, m}$ using precomputed $\bm{\mu}_m$ equations~\eqref{eqn:pmm}~\&~\eqref{eqn:pm1m} \\
			}
			\FOR{ every  $\ell = m+2, ..., \ell_{max}$}
				\STATE {
					$\circ$ compute ${\cal P}_{\ell m}$ via the 2-point recurrence, given precomputed $\bm{\beta}_{\ell m}$, equation~\eqref{eqn:assLegRec} \\
					$\circ$ update $\bm{\Delta}^{A}_m\l( r\r)$, equation~\eqref{eqn:DeltaDef} \\
				}
			\ENDFOR{ ($\ell$)}
		    \ENDFOR{ ($r$)}
		\ENDFOR{ ($m$}
		\medskip
		\STATE{{\sc global communication}}\\
		\STATE{ \hspace{0.2cm} $\circ$ $\l\{\bm{\Delta}^{A}_{m}\l(r\r),\; m\in {\cal M}_i,\; \hbox{{\rm \underline{all} $r\in{\cal R}$}}\r\} {\hbox{ {\tt \scriptsize MPI\_Alltoallv}}\atop\Longleftrightarrow} \l\{\bm{\Delta}^{A}_{m}\l(r\r),\; r\in {\cal R}_{i},\; \hbox{{\rm \underline{all} $m\in{\cal M}$}}\r\}$ }
		\medskip
		\STATE{\sc step 3 - $\bm{s}_{n}$ calculation}\\
		\FOR{every ring $r \in {\cal R}_i$}
			\STATE{ $\circ$ via FFT calculate $\bm{s}\l(r, \phi\r)$ for  all samples $\phi$ of ring $r$,  given pre-computed $\bm{\Delta}^{A}_m\l(r\r)$ for  \underline{all} $m$. }
		\ENDFOR{($r$)}
		\medskip
		\RETURN {$\bm{s}_{n}$ for all $\l(r\in {\cal R}_i\r)$}
	\end{algorithmic}
	\label{algo:Generallalm2mapS2HATParallel}
\end{algorithm*}
The operations listed there are to be performed by each of the
$n_{proc}$ MPI processes involved in the computation. Like its serial
counterpart, the parallel algorithm can be also split into
steps. First, we precompute the starting values of the Legendre
recurrence $\bm{\mu}_{m}$ \eqref{eqn:pmm}, but only results for $m \in
{\cal M}_i$ are preserved in memory. In next step, we calculate the
functions $\bm{\Delta}^{A}_m$ using equation~\eqref{eqn:DeltaDef} for
{\em every} ring $r$ of the sphere, but only for $m$ values in ${\cal
  M}_i$.  Once this is done, a global (with respect to the number of
processes involved in the computation) data exchange is performed so
that at the end each process has in its memory the values of
$\bm{\Delta}^{A}_m(r)$ computed only for rings $r\in{\cal R}_{i}$ and
for {\em all} $m$ values. Finally, via FFT we synthesise the partial
results as in equation~\eqref{eqn:Delta2map}. From the point of view
of the framework, steps 1 and 2 constitute the first stage of the
computation, followed by the global communication, and the second
computation involving here a calculation of the FFT. As mentioned
earlier, though the communication is an inherent part of the
framework, the two computation stages can be adapted as
needed. Hereafter, we adhere to the standard SHT algorithm as used in
Algorithm~\ref{algo:Generallalm2mapS2HATParallel} .

\subsubsection{Performance analysis}

\label{sec:perfan}

\paragraph{Operation count.}

We note that the scalability of all the major data products can be
ensured by adopting an appropriate distribution of the harmonic
coefficients and the map.  This stems from the fact that apart of the
input and output objects, i.e., the maps and the harmonic coefficients
themselves, the volumes of which are inversely proportional to the
number of the MPI processes by construction.  The largest data objects
derived as intermediate products are the arrays $\Delta_A$ and
$\Delta_S$, whose sizes are ${\cal O}({\cal R}_N\,m_{max}/n_{proc})$,
and thus again decreasing inversely proportionally with $n_{procs}$.
These objects are computed during the first computation stage, then
redistributed during the global communication step, and used as the
input for the second stage of the computation.  Their total volume, as
integrated over the number of processes, therefore also determines the
total volume of the communication, as discussed in the next
section. As these three objects are indeed the biggest memory
consumers in the routines, the overall memory usage scales as
$(n_{pix}+2\,m_{max}\,\ell_{max}+{\cal R}_N\,m_{max})/n_{procs}$.  For
typical pixelization schemes and full sky maps analysed at its
sampling limit, we have $n_{pix} \sim \ell_{max}^2$, $\ell_{max}
\simeq m_{max}$, and ${\cal R}_N \sim n_{pix}^{1/2}$, and therefore
each of these three data objects contribute nearly evenly to the total
memory consumption of the routines. Consequently, the actual high
memory water mark is typically $\sim50$\% higher than the volume of
the input and output data. We note that if all three parameters,
$\ell_{max}, \ m_{max}, {\cal R}_N$, can have arbitrary values, then
reusing the memory, and, in particular an {\em in-place}
implementation of the transforms is not straightforward, as it may not
be easy or possible to fit any of these data objects in the space
assigned for other data objects. For this reason, such options are not
implemented in the software discussed here.  Given the assumed data
layout, also the number of flops required for the evaluation of FFTs
and associated Legendre functions scales inversely proportionally to
the number of the processes. The only exception is related to the cost
of the initialisation of the recurrence, equation~(\ref{eqn:pmm}),
which is performed redundantly by each MPI process to avoid extra
communication. As each process receives at least one value of $m$ on
order of $m_{max}$, it has to perform at least ${\cal O}(m_{max)}$
operations as part of the pre-computation, equation~\eqref{eqn:pmm}. A
summary of the flops required on each step of the parallel SHT
algorithms is presented in Table~\ref{tab:flops}, where we have
assumed here that the subsets ${\cal R}_{i}$ and ${\cal M}_{i}$
contain $ \simeq {\cal R}_{N}/n_{proc}$ and $\simeq m_{max}/n_{proc}$
elements, respectively.
\begin{table}[h]
	\begin{center}
		\begin{tabular}{llc} 
			&  & \emph{Flops} \\ 
			\hline\hline
			\medskip
			{\sc Pre-computation}	& {\sc step \{2,1\} } &  $O\left(m_{max}\right)$ \\
			\medskip
			{\sc Recurrence}	& {\sc step \{3,2\} } & $O\left(({\displaystyle {\cal R}_{N}\ell_{max} \frac{m_{max}}{n_{proc}}}\right)$\\
			\medskip
			{\sc FFTs }	& {\sc step \{1,3\} } & $O\left({\displaystyle \frac{ {\cal R_{N}}} {n_{proc}}\;m_{max}\log m_{max}}\right)$\\ \hline
		\end{tabular}
	\end{center}
\caption{Estimates of floating point operations required in the parallel SHT algorithm.}\label{tab:flops}
\end{table}

\paragraph{Communication cost.}

The overall performance of the algorithms depends also on the
communication cost. The data exchange is performed by using a single
collective operation of the type all-to-all, in which each process
sends a distinct message to every other process. Hereafter, we
assume that the time needed to send a message of $S_{msg}$ bytes
between any two MPI process can be estimated by $\alpha + \beta
S_{msg}$, where $\alpha$ is the latency for sending a message and
$\beta$ is the inverse of the bandwidth.  Different MPI libraries use
different specific algorithms to implement such a communication, which
moreover may be different for different sizes of the message.  For
concreteness, we base our analysis on MPICH~\cite{mpich}, a widely
used MPI library. Similar considerations can be performed for other
libraries.  For short messages ($\leq 256$ kilobytes) where the
latency is an issue, the all-to-all communication in
MPICH~\cite{collective_mpi} uses the index algorithm by Bruck et
al.~\cite{bruck}. This algorithm features a trade-off between the
startup time and the data transfer time, and while taking only $ \log
n_{proc}$ steps, it communicates data volumes potentially as big as
$S_{msg}(n_{proc}/2) \log n_{proc}$.
For long messages and an even number of processes, MPICH switches to a
pairwise-exchange algorithm, which performs series of exchanges among
pairs of processes.  In the case of uniform distributions of the rings
$r$ and multipoles $m$ between processes, the size of the message to
be exchanged between each pair of processes is given as,
\begin{equation}\label{eq:msg_size}
	S_{msg} := {\cal R}_{N}\frac{m_{max}}{n_{proc}}n_{\mathbb C},
\end{equation}
where $n_{\mathbb C}$ is the size in bytes of a complex number
representation (usually $n_{\mathbb C} = 16$).  Using this expression
and all the assumptions listed above, the following formula estimates
the total time required for communication in our parallel SHT
algorithm,
\begin{equation}
	T_{comm} = 
	\begin{dcases*}
	        \alpha \log n_{proc} + \beta\,S_{msg}\frac{n_{proc}}{2}\log n_{proc}, & when $S_{msg}\leq$ 256\,kB, \\
	        \alpha \l(n_{proc} - 1\r) + \beta\,S_{msg}\,\l(n_{proc}-1\r), & when $S_{msg}>$ 256\,kB.
	        \end{dcases*}.
	        \label{eqn:comm}
\end{equation}

\begin{figure}[ht]
\centering

\includegraphics[width=0.325\textwidth]{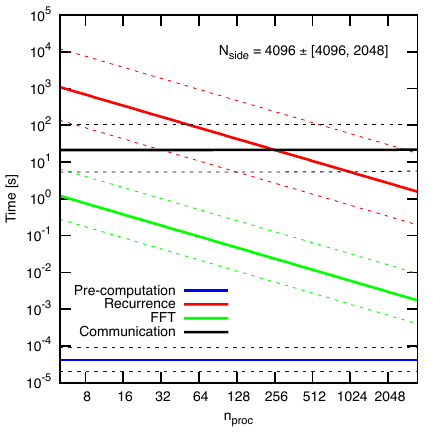}
\includegraphics[width=0.325\textwidth]{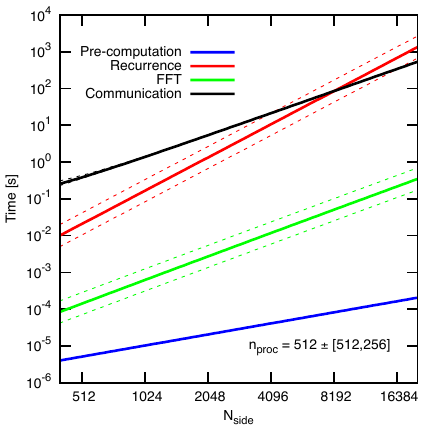}
\includegraphics[width=0.325\textwidth]{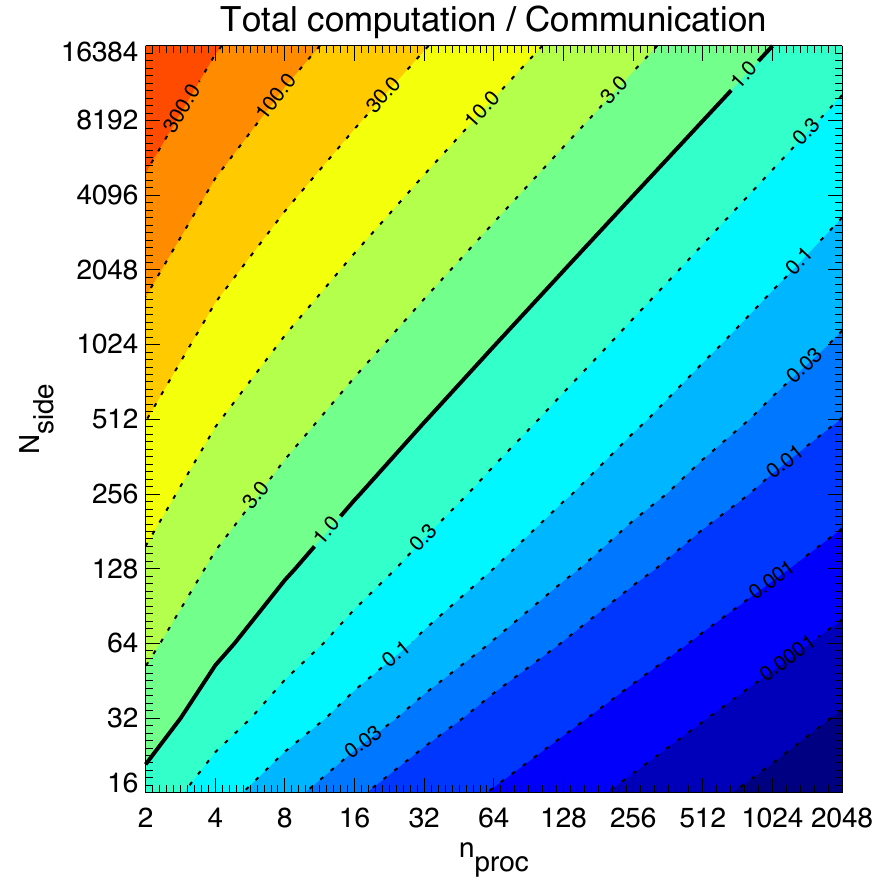}

\caption{Two plots on the left show theoretical runtimes (logarithmic
  scale) as function of the number of processes for a fixed size of
  the problem, leftmost panel, and as a function of the
  resolution/size of the problem for a fixed number of processes,
  middle. The sizes of the problem shown in the leftmost panel with
  thick, solid lines, correspond to $\simeq 2\;10^8$ pixels and a
  similar number of the harmonic modes, corresponding to the HEALPix
  parameter value $N_{side} = 4096$.  Thin, dashed lines show cases
  with $4$ times fewer (more) pixels, lower and upper lines
  respectively.  In the middle panel the number of processes is
  assumed to be $n_{proc}=512$, thick, solid line, or 256 (1024), as
  shown by thin dashed lines.  The contours in the rightmost panel
  display a ratio of the total computation to communication times
  shown as a function of the number of process and the size of the
  problem.  }\label{fig:theoretical_performance}
\end{figure}

In Figure~\ref{fig:theoretical_performance} we illustrate theoretical
runtimes depending on the number of the employed processes (leftmost
panel) and on the size of the problem (middle panel).  The transform
parameters have been set assuming a standard full sky, fully resolved
case, and the HEALPix pixelization, i.e., $\ell_{max} = m_{max} = 2 \,
N_{side}, \ {\cal R}_N = 4\,N_{side}-1, \ n_{pix} = 12 \,
N_{side}^2$. Here, $N_{side}$ is the HEALPix resolution parameter (for
more details see beginning of section~\ref{sec:experiments} and
\cite{Gorski_etal_2005}).  For the hardware parameters, in the
estimation of the communication cost we have used,
following~\cite{predigting_run_times,passinggraph}, $\alpha =10^{-5}$,
$\beta=10^{-9}$, while for the cost of calculations we have assumed
that each MPI process attains the effective performance of $10$ Gflops
per second, resulting in a time of $\gamma=10^{-10}$ seconds per
flop. The choice we have made is meant to be only indicative, yet at
its face value the latter number is more typical of the MPI processes
being nodes of multi-cores than just single cores, in particular given
the usual efficiency of such calculations, which is usually on the
order of $10-20$\%. If less efficient MPI processes are used, then the
solid curves in Figure~\ref{fig:theoretical_performance} corresponding
to the computation times should be shifted upwards by a relevant
factor.
 
The dominant calculation is the recurrence needed to compute the
associated Legendre functions in agreement with our earlier numerical
result, Figure~\ref{fig:pie_plot}. The computation scales nearly
perfectly with the number of the used processes, decaying as $\propto
1/n_{proc}$ while increasing with the growing size of the problem,
$\propto \ell_{max}^2 \propto N_{side}^2$. At the same time, the
communication cost is seen to be independent of the number of
processes. This is because for the numbers of processes considered
here, the size of the message exchanged between any two processes is
always large and the communication time is thus described by the
second term of the second equation of the set of
equations~\eqref{eqn:comm}. Given that the single message size,
$S_{msgs}$, decreases linearly with the number of processes, the total
communication time does not depend on it. The immediate consequence of
these two observations is that both these times will be found
comparable if a sufficiently large number of processes is used. A
number of processes at which such a transition takes place will depend
on the constants entering in the calculation of
equations~\eqref{eqn:comm} and the size of the problem.  This
dependence of the critical number of processes on the size of the
problem is shown in the rightmost panel of
Figure~\ref{fig:theoretical_performance}, with thick solid contour
labelled $1.0$.  Clearly, from the perspective of the transform
runtimes, there is no extra gain which could be obtained from using
more processes than given by their critical value.  In practice, the
number of processes may need to be fixed by memory rather than by
computational efficiency considerations. As the memory consumption
depends on the problem size, i.e., $n_{pix}$, in the same way as the
communication volume, by increasing the number of processes in unison
with the size of the problem we can ensure that the transform runs
efficiently, i.e., the fraction of the total execution time spent on
communicating will not change, and that it will always fit within the
machines memory.

We note that some improvement in the overall cost of the communication
could be expected by employing a non-blocking communication pattern
applied to smaller data objects, for example single rows of the arrays
$\Delta_A$ and $\Delta_S$ corresponding to a single ring or $m$ value,
successively and immediately after they are computed. Such an approach
could extend the parameter space in which the computation time is
dominant and the overall scaling favorable, by overlapping
communication with some computation. We expect this to be beneficial
only for the inverse transform as for the direct one the computation
preceding the communication, and involving the FFTs, is subdominant
with respect to both the communication and the recurrence time in the
regime where these both are becoming comparable.  Therefore, the gain
from overlapping the FFT calculation with the communication could be
at the best minor. Moreover, even for the inverse transform the gain
would be rather limited as long as the same, large, total volume of
data, is required to be communicated and the communication time will
eventually supersede that of the calculation at some concurrency. We
note that the communication volume could be decreased, and thus again
the parameter space with the computation dominance extended, if
redundant data distributions and redundant calculations are allowed
for in the spirit of the currently popular {\em communication
  avoiding} algorithms, e.g., \cite{grigori2008communication}. Though
these kinds of approaches are clearly of interest they are outside the
scope of this work and are left here for future research.
 
We also note that the conclusions may disfavour algorithms which
require big memory overhead and/or need to communicate more. Such
algorithms will tend to require a larger number of processes to be
used for a given size of the problem, and thus will quickly get into
the bad scaling regime. Their application domain may be therefore
rather limited.  We emphasise that this general conclusion should be
reevaluated case-by-case for any specific algorithm. Likewise, it is
clear that the regime of the scalability of the standard SHT algorithm
could be extended by decreasing the communication volume. Given that
the communicated objects are dense, the latter would typically require
some lossy compression techniques and therefore we do not consider
them here. Instead we focus our attention on accelerating intra-node
calculations. This could improve the transforms runtimes for a small
number of processes.  However if successful, unavoidably it will also
result in lowering the critical values for the processes numbers,
leading to losing the scalability earlier. Nevertheless, the overall
performance of these improved transforms would never be inferior to
that of the standard ones whatever the number of processes.

\subsection{Exploiting intra-node parallelism}\label{sub:acclerating} 

In this section we consider two extensions of the basic parallel
algorithm, Algorithm~\ref{algo:Generallalm2mapS2HATParallel}, each
going beyond its MPI-only paradigm. Our focus hereafter is therefore
on the computational tasks as performed by each of the MPI processes
within the top-level framework outlined earlier. Guided by the
theoretical models from the previous section, we expect that any gain
in performance of a single MPI process will translate directly into
analogous gain in the overall performance of the transforms, at least
as long as the communication time is subdominant, that is in a
relatively broad regime of cases of practical importance. Hereafter we
therefore develop implementations suitable for multi-core processors
and GPUs and discuss the benefits and trade-offs they imply.

As highlighted in Figure~\ref{fig:theoretical_performance}, the
recurrence used to compute the associated Legendre functions takes by
far a dominant fraction of the overall runtime of the three
computational steps.  Consequently, we will pay special attention to
accelerating this part of the algorithm, while we keep on using
standard libraries to compute the FFTs. The latter will be shown to
deliver sufficient performance given our goals here.  The associated
Legendre function recurrence involves three nested loops and we will
particularly consider advantages and disadvantages of their different
ordering from the perspective of attaining the best possible
performance. We note the recurrence is involved on the stages when
either the harmonic coefficients $\bm{a}_{\ell m}$ are computed given
the already available $\bm{\Delta}_m^A$ in the case of the direct
transform, or when the computation of $\bm{\Delta}_m^S$ is performed
given the input coefficients $\bm{a}_{\ell m}$. In both cases, each
shared-memory node stores information concerning all observed rings
and a subset of $m$ values as defined by ${\cal M}_i$ and as
distributed on the memory-distributed level of parallelisation.
Hereafter, for simplicity we will refer to the algorithms for the
direct and inverse spherical transforms by the core names of their
corresponding routines, \map2alm and \alm2map respectively.

\subsubsection{Multithreaded version for multi-core processors.}\label{sub:multithreading_alm2mapmt} 

Since the largest and the fastest computers in the world today employ
both shared and distributed memory architectures, their potential can
not be fully exploited by software which uses only distributed memory
approaches such as MPI.  To overcome this problem, hybrid approaches,
mixing MPI and shared memory techniques such as multithreading have
been proposed. In this spirit, we introduce a second level of
parallelism based on the multithreaded approach in the computation of
SHT.

We note that the three nested loops which we need to perform to
calculate the transforms are in principle interchangeable.  In
practice, the loop over $\ell$ has to be the innermost one in order to
avoid significant memory overhead or repetitive re-calculations of the
same objects.  Out of the two remaining loops, selecting the loop over
$m$ as the outermost, as in the MPI-only case from
Algorithm~\ref{algo:Generallalm2mapS2HATParallel}, allows precomputing
$\beta_{\ell m}$, equation~\eqref{eqn:betaDef}, ${\cal P}_{m m}$,
equation~\eqref{eqn:pmm}, and ${\cal P}_{m+1, m}$,
equation~\eqref{eqn:pm1m}, only once for each ring. This also requires
extra storage  to store $\beta_{\ell m}$ for a given $m$,  which is however only on the 
order of ${\cal O}\l(\ell_{max}\r)$, as well as the intermediate results, i.e.,
$\bm{\Delta}_m^{A/S}$, which  require additional storage on the order of
${\cal O}\l(\ell_{max}\,{\cal R}_N\r)$. This latter estimate is comparable to the
size of the input and output, and therefore not negligible in the
memory budget of the calculation. Nevertheless, given that it
decreases with the number of MPI process and given the sizes of memory
banks of modern supercomputers, tens of Gigabytes per shared memory
node, this can be managed in practice straightforwardly, as has been
indeed demonstrated earlier in the case of the MPI-only routines.
This choice of the loop ordering is therefore efficient
from both computation and memory points of view.  
Consequently, in the multithreaded versions of Algorithms $1$ and $2$
the outermost loop, parallelised using OpenMP, is the loop over $m$.
The set of $m$ values, assigned to each MPI process $i$,
i.e., ${\cal M}_{i}$, is then evenly divided into a number of subsets equal
to the number of threads mapped onto each physical core available on a
given multi-core processor. We denote this subset of $m$ values as
${\cal M}^{\text{T}_i}_j$, where subscript $j$ corresponds to the
thread number. In such a way, each physical core executes one thread
concurrently and calculates intermediate results of SHT for its local
(in respect to the shared memory) subset of $m$. Algorithm
\ref{algo:alm2mapS2HATParallelMT} describes \text{\sc step 1} of
Algorithm~\ref{algo:Generallalm2mapS2HATParallel} in its multithreaded
version.
\begin{algorithm}
	\caption{{\sc $\bm{\Delta}^{A}_m$ calculation per thread}}
	\begin{algorithmic} 
		\REQUIRE {all $m\in{\cal M}^{\text{T}_{j}}_{i}\;\Rightarrow \bigcup \limits^{\text{Nth}}_{j=1}{\cal M}_{i}^{\text{T}_j}={\cal M}_{i}$ }
		\FOR{all $m \in {\cal M}^{\text{T}_j}_i$}		
			\STATE {
				$\circ$ initialise ${\cal P}_{m\, m}$ and ${\cal P}_{m\, m+1}$, equations~\eqref{eqn:pmm} \\
                                     $\circ$ precompute $\beta_{\ell m}$,~equation~\eqref{eqn:betaDef} \\				
			}
			\FOR{ \underline{every} ring $r$}
				\FOR{ every  $\ell = m+2, ..., \ell_{max}$}
					\STATE {
						$\circ$ compute ${\cal P}_{\ell m}$ via $\bm{\beta}_{\ell m}$, equation~\eqref{eqn:assLegRec} \\
						$\circ$ update $\bm{\Delta}^{A}_m\l( r\r)$ in the shared memory, equation~\eqref{eqn:DeltaDef} \\
						}
				\ENDFOR{ ($\ell$)}
			\ENDFOR{($r$)}
		\ENDFOR{($m$)}
	\end{algorithmic}
	\label{algo:alm2mapS2HATParallelMT}
\end{algorithm}


\subsubsection*{Load balance between active threads.}

As already noted for the MPI-only cases, the loop ordering adopted
here does not ensure proper load balance between different threads.
Nevertheless, it is flexible enough to leave room allowing for
appropriate tuning. As before, potential imbalance of the
computational load is due to the different number of steps which need
to be performed during the recursive calculation of the associated
Legendre functions, and which is equal to $\ell_{max}-m+1$.  Therefore
it explicitly depends on $m$. To avoid this problem, a care needs to
be taken while defining the specific subsets of $m$ values, which are
to be assigned to each thread, ${\cal
  M}^{\text{T}_j}_{i}$. \begin{figure}[ht] \centering
  \includegraphics[width=0.95\textwidth]{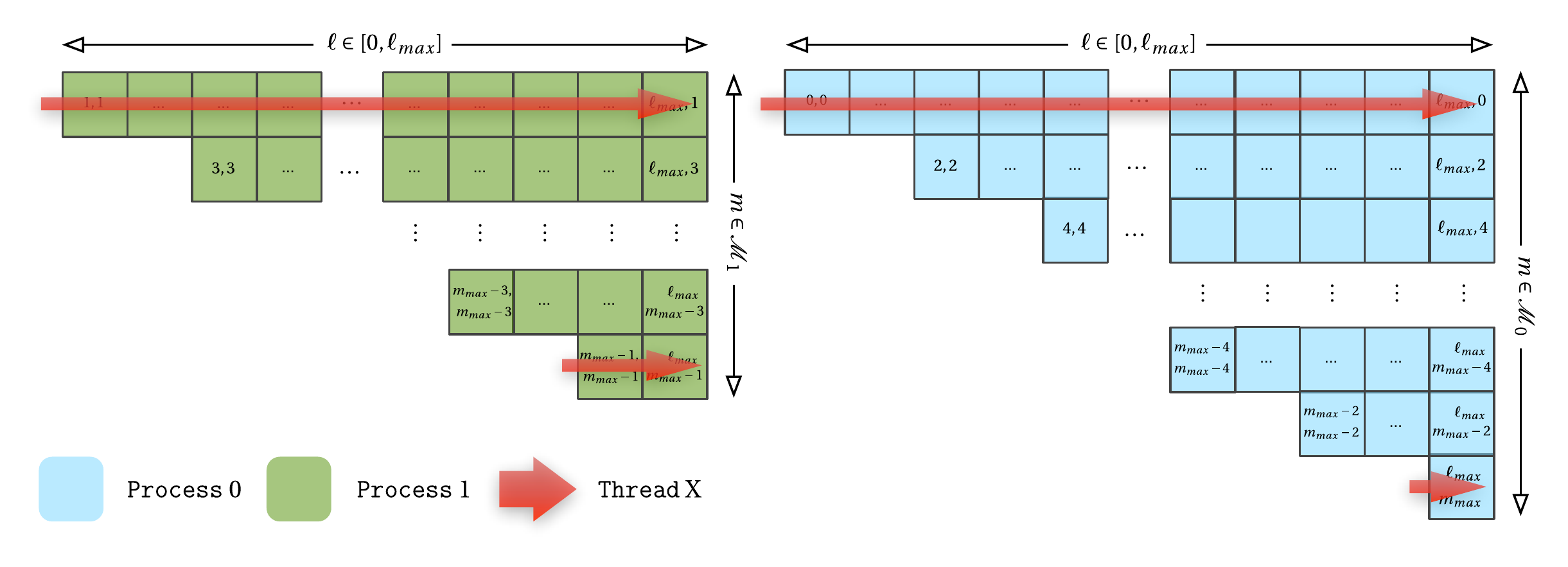}
\caption{Example of two "min-max" pairs of $m$ values assigned to {\cal Thread} $X$ on Process 1 (left) and 0 (right) respectively. The read arrow represents the direction and the length of the $\ell$-recurrences (equation~\eqref{eqn:pm1m}) evaluated by a thread for a given "min-max" pair. }\label{fig:mt_load_balance}
\end{figure}
Hereafter, we do so by replicating the same approach as used earlier
to balance the workload on the distributed memory level, and generate
the subsets, $M^{\text{T}_j}_{i}$, by taking values from the set
${\cal M}_{i}$ in the "max-min" fashion i.e., sequentially we pick
from ${\cal M}_{i}$ the maximal and minimal values as available at any
given time and assign them to a chosen thread, while at the same time
we try to keep the number of such pairs per thread nearly the same for
every thread.  Example of such a mapping is depicted in
Figure~\ref{fig:mt_load_balance}.

We note that if such a distribution of $m$ modes is done consistently
on both distributed- and shared- memory levels, it will not only
result in balancing the workload within the single node but also
across all the nodes involved on the MPI level, ensuring global load
balance. That such balance indeed follows can be seen immediately by
noticing that on the distributed memory level each set ${\cal M}_i$
assigned to one of the nodes consists of pairs of $m$ values, $\{m',
m''\}$, picked as a min-max pair and the sum of which is constant and
given by $m'+m''=m_{max}$. Consequently, the same numerical cost is
incurred in the computation of the associated Legendre functions for
each pair, as the latter is completely determined by the number of the
recursion steps, which is given by $\ell_{max}-m'+1+\ell_{max}-m''+1 =
2\ell_{max} -m_{max}+2$ and therefore independent of both $m'$ and
$m''$.  At the same min-max principle applied on the intra-node level
will assign each such pairs to one of the threads and therefore lead
to a near perfect load balancing as long as the number of pairs
assigned per thread is nearly the same for all the threads. Of course,
departures from the well-balanced load may be expected whenever the
number of $m$ values assigned per node or per thread becomes small.

\subsubsection{SHT on GPGPUs.}\label{sub:sht_on_gpu} 

A CUDA (Compute Unified Device Architecture) device is a chip
consisting of a number of Streaming Multiprocessors (SM), each
containing hundreds of simple cores (typically $512$), called
Streaming Processors (SP), which execute instructions sequentially.
The Streaming Multiprocessors perform instructions in a SIMD (Single
Instruction Multiple Data) fashion and can maintain hundreds of
active threds, referred to as a thread block hereafter, which are
launched in batches of typically $32$ threads, called warps. The warps
are executed sequentially in the cyclic order, and this helps to hide
the memory latency between the local memory of SMs and the global
memory of the GPU. The former memory is shared within the SMs, it is
fast but limited.  For example it is up to $48$KB in size for the
latest Nvidia cards. There is no synchronisation between different SMs
and they can communicate only through relatively slow global memory.

\begin{figure}[ht]
  \centering
  \includegraphics[width=0.95\textwidth]{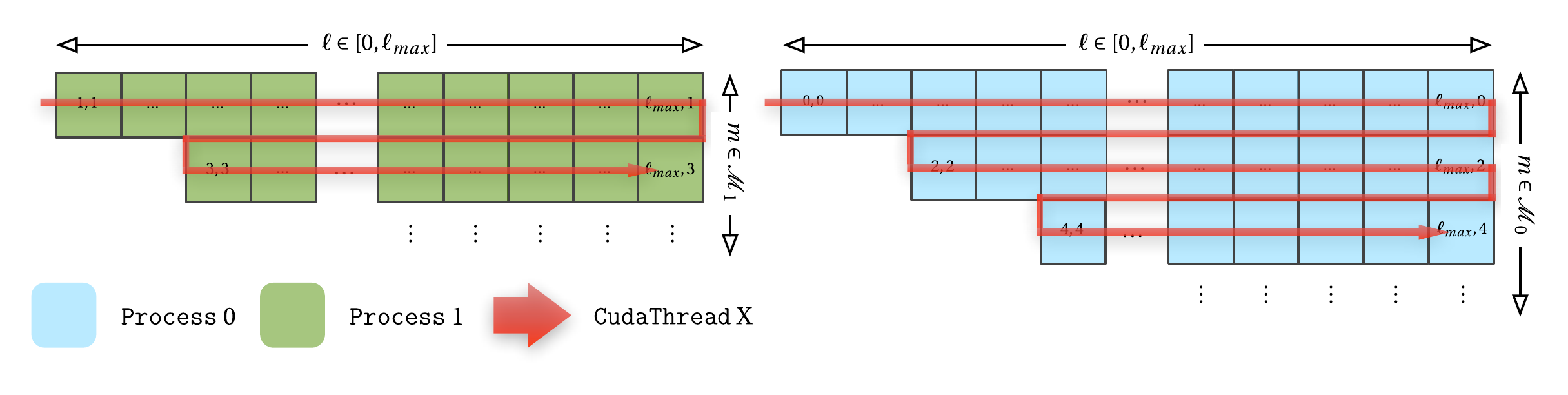}\caption{Schematic
    view of the iterations of the recurrence with respect to $\ell$
    (equation~\ref{eqn:pm1m}) evaluated by each thread on CUDA device
    assigned to two different GPUs, referred to as Process $1$ and $0$
    respectively.}\label{fig:cuda_alm_access}
\end{figure}

The fact that all the active threads of the GPU execute the same
instruction set suggests that the most suitable approach for this
device is to choose the loop over the rings, $r$, as the outermost
one, and to parallelise the calculation with respect to it.  This is
contrary to the multi-node implementation, as described earlier, which
is based on setting the loop over $m$ as the outermost one. The latter
choice has the advantage of optimising the number of numerical
operations, such as those involved in the calculation of the \blm{}
object, which is independent on $r$ and can be computed for each value
of $m$ as set by the outermost loop and used for all values of $r$ in
the loop over $r$. The downside of this approach is that if the
outermost loop is parallelised and different values of $m$ assigned to
different threads, then the innermost loop over $\ell$ number will
require a different number of passes depending on the specific value
of $m$. Consequently, the parallelisation on the finest level, i.e.,
within each warp, will be lost.  Setting the loop over $r$ as the
outermost not only avoids this problem but also permits using more
active threads, as at this stage of the calculation each GPU will
store in its memory the object $\Delta_m( r)$ distributed between the
GPUs with respect to $m$, with each GPU storing only those of its
entries which correspond to the subset ${\cal M}_i$ of all values of
$m$ but for all values of $r$ $(\in {\cal R})$.  These advantages seem
to offset the increased computational load as involved in the
re-calculation of \blm{}, equation~\eqref{eqn:pmm}, and we use
therefore this approach to implement the SHT algorithms on GPUs.  The
respective algorithm then involves, first, dividing the set of all
rings into disjoint subsets, and subsequently assigning them to
different threads of every GPU used in the computation. The threads
perform then the calculation for all the rings assigned to it as well
as all values of $m$ from the subset, ${\cal M}_i$, as assigned to
this GPU (and therefore common to all its threads),
Figure~\ref{fig:cuda_alm_access}.  The overall load balancing of the
calculation is ensured if the number of rings per thread is the same
for all threads, and the values of $m$ are distributed between the
GPUs in the same way as in the case of the multi-node algorithm.  We
note that a similar approach has been adopted in \cite{hupca2012} in
the case of their \alm2map\ routine, and we will hereafter use their
solution to optimise the calculation of \blm{}.  This approach
achieves the best performance if one ring is assigned per GPU thread,
as then the involved recalculation can be completely avoided.

Setting the loop over the rings  as the outermost one enables two completely different approaches for the memory management in the direct and inverse SHT. We describe them in the following two paragraphs.

\subsubsection*{\map2alm.}

Mapping the rings onto a big number of threads facilitates the
parallelisation of the SHT algorithm on GPUs, as it permits each
active thread to perform exactly the same number and type of
operations but with different values. However, in the case of the
direct transform each thread calculates a contribution to all non-zero
values of the harmonic coefficient matrix $\bm{a}_{\ell,m}$ for $m\in
{\cal M}_i$ and $\ell \in [0, \dots, \ell_{max}]$, and coming from a
subset of all rings as assigned to this thread. The calculation of the
final result therefore requires an all-reduce operation involving all
the active threads within the GPU. As on CUDA the active threads can
communicate only within the thread block, the task is not completely
straightforward.

Algorithm~\ref{alg:cuda_map2alm} outlines {\sc step 3} of
Algorithm~\ref{algo:map2almBasic} adapted to CUDA. The set of values
of $m$ to be processed by a selected GPU is given by the set ${\cal
  M}_i$, assigned to each GPU on the distributed-memory level. To take
advantage of the fast shared memory, and to simultaneously avoid the
high latency memory of the device, we first calculate the values of
\blm{} vectors in segments, which can fit in the shared memory i.e.,
all threads within a block in a "collaborative" way compute in advance
\blm{} values and store them in shared memory. Next, these precomputed
values are used to evaluate an associated Legendre function ${\cal
  P}_{\ell m}$, followed by each active thread fetching from global
memory the precomputed sum $\bm{\Delta}^{S}_m\l( r\r)$ and performing
the final multiplication from equation~\eqref{eqn:almdef}. Once it is
done, we sum all these partial results within a given block of threads
using an adapted\footnote{We omitted a part of the code, which writes
  the partial result to global memory} version of the parallel
reduction primitive from CUDA SDK. We note that for typical values of
$\ell_{max}$, the memory needed to store all the harmonic coefficients
$\bm{a}_{\ell m}$ exceeds the shared memory available in each SM. They
will have to be therefore communicated to the global memory in batches
with sizes close to that of the shared memory.  Consequently at any
time all the thread blocks will send the data concerning the same
subset of the harmonic coefficients and do so multiple times.  This
implies that as many as ${\tt NBT}$ threads, where ${\tt NBT}$ is the
number of blocks of threads, will try to update the values of the same
variables stored in the same space of the global memory. This leads to
a \emph{race condition problem}. The race conditions can be resolved
in CUDA by using atomic operations (e.g., via CUDA {\tt atomicAdd}
function), which are capable of reading, modifying, and writing a
value back to the shared or global memory without an interference from
any other thread, which guarantees that the race condition will not
occur. However, if two threads perform an atomic operation at the same
memory address at the same time (which may be the case in our
algorithm), those operations will be serialized i.e., in this
situation each atomic operation will be done one at a time, but
execution will not progress past the atomic operation until all the
threads have completed it. Consequently, atomic operations may cause
dramatic degrade of the performance which additionally strongly
depends on the hardware\footnote{For instance, Nvidia
  reports~\cite{FERMI:Nvidia} that there is a substantial difference
  between atomic operation performance between CUDA with compute
  capability 1.1 (where it first appeared) and more recent Fermi
  devices (compute capability 2.x).}. Alternately, we can avoid the
race conditions and thus, execution of the expensive atomic functions
by increasing granularity of the recurrence step. In this scenario we
loop over $m$ values on CPU and execute as many \emph{small} CUDA
kernels as elements in ${\cal M}_i$, which for given values $m$ and
$\bm{\mu}_m$, evaluate ${\tt NBT}$ reduced (per block of threads)
results $a^{\star}_{\ell m}$, which are next copied from the device to
the host memory and summed together on CPU. This variant of our
algorithm corresponds to the state {\tt HYBRID\_EVALUATION = true}, in
Algorithm~\ref{alg:cuda_map2alm}.  This clearly could be the only
option if the atomic operations are not supported on the GPUs used for
the calculation. We note however that even if they are supported, this
latter approach may prove to be more efficient, as it indeed was the
case for the GPU used in this work. For this reason, all the
experimental results discussed in the next section are calculated with
the reduction performed on a CPU.

\begin{algorithm}
	\caption{{\sc \aalm{} calculation on CPU\&GPU }}
	\begin{algorithmic}
		\REQUIRE {$\bm{\Delta}^{S}_m$ \& $\bm{\mu}_m$ vectors in the GPU global memory for $m\in{\cal M}_i$}
		\FOR{ \underline{every} ring $r$ assign to one GPU thread} 
			\FOR{every $m \in {\cal M}_i$}
				\FOR{ every  $\ell = m+2, ..., \ell_{max}$}
					\STATE{ $\bullet$ use precomputed or, if needed, precompute in parallel a segment of \blm{}, equation~\eqref{eqn:betaDef}; }
					\STATE{ $\bullet$ compute ${\cal P}_{\ell m}$ via equation~(\ref{eqn:assLegRec});	}
					\STATE{ $\bullet$ evaluate: ${a}^{r}_{\ell m} = \bm{\Delta}^{S}_m\l( r\r) {\cal P}_{\ell m}(r)$}
					\STATE{ $\bullet$ sum all partial results ${a}^{r}_{\ell m}$ within a block of threads 
						\begin{equation}\label{eq:cuda_partial_alm}
							{a}^{\star}_{\ell m} = \sum^{\tt NTB }_{i = 0} \l({a}^{r}_{\ell m}\r)_{i} 
						\end{equation}
						and store the result in the shared memory.
				 	}
					\IF {{the shared memory is filled}}
					\IF {{\tt HYBRID\_EVALUATION}}
					 	\STATE{ $\bullet$ for each block of threads with different id = {\tt BID}, write $\bm{a}^{\star}_{\ell m}$ in the device global memory}
						\begin{equation*}
							\bm{a}^{GPU}_{\ell m}[{\tt BID}] = {a}^{\star}_{\ell m} 
						\end{equation*}
						\STATE{ $\bullet$ copy partial results $\bm{a}^{GPU}_{\ell m}[\ldots]$ from  the device to the host memory and sum together using CPU}
						\FOR{$i = 0 \to {\tt NBT}$} 
						\STATE {$\bm{a}_{\ell m} = \bm{a}_{\ell m} + {a}^{\tt GPU}_{\ell m}[i]$ }
						\ENDFOR
					\ELSE
						\STATE{ $\bullet$ update $\bm{a}_{\ell m}$ in the global memory via an atomic addition: $\bm{a}_{\ell m} = \bm{a}_{\ell m} + {a}^{\star}_{\ell m}$ }
					\ENDIF
					\ENDIF

					\medskip  
				\ENDFOR{ ($\ell$)}
			\ENDFOR{($m$)}
		\ENDFOR{($r$)}
		\medskip
		\RETURN{$\bm{a}_{\ell,m}$: $ m\in {\cal M}_i$ and ${\ell \in [2,\dots,\ell_{max}}]$ }
	\end{algorithmic}
	\label{alg:cuda_map2alm} 
\end{algorithm}

\subsubsection*{\alm2map.}

The inverse SHT also requires a reduction of the partial results, but contrary to the direct transform, the summation is done over $\ell$ as shown by equation~\eqref{eqn:DeltaDef}. With the rings assigned to single threads, this leads to  a rather straightforward implementation. Algorithm~\ref{algo:alm2mapS2HATParallelCUDA} outlines the \alm2map algorithm devised for GPUs and it is based on the version detailed in \cite{hupca2012}. 
\begin{algorithm}
	\caption{{\sc $\bm{\Delta}^{A}_m$ calculation on GPU}}
	\begin{algorithmic}
		\REQUIRE {$\bm{a}_{\ell,m}\in\mathbb{C}^{m_{max}\times\ell_{max}}$ matrix in GPU global memory} 
		\FOR{ \underline{every} ring $r$ assign to one GPU thread} 
			\FOR{$m \in {\cal M}_i$}
				\FOR{ every  $\ell = m+2, ..., \ell_{max}$}
					\STATE{ $\bullet$ use fetched or, if needed, fetch in parallel a segment of \aalm{} data; 	} 
					\STATE{ $\bullet$ use precomputed or, if needed, precompute in parallel a segment of \blm{}, equation~(\ref{eqn:betaDef}); }
					\STATE{ $\bullet$ compute ${\cal P}_{\ell m}$ via equation~(\ref{eqn:assLegRec});	}
					\STATE{ $\bullet$ update $\bm{\Delta}^{A}_m\l( r\r)$ for a given prefetched $\bm{a}_{\ell m}$ and computed ${\cal P}_{\ell m}$, equation~\eqref{eqn:DeltaDef}; }
				\ENDFOR{ ($\ell$)}
				\STATE{ $\bullet$ save final $\bm{\Delta}^{A}_m\l( r\r)$ in global memory }
			\ENDFOR{($m$)}
		\ENDFOR{($r$)}
		\medskip
		\RETURN{vector of partial results: $\bm{\Delta}^{A}_m\in\mathbb{C}$ }
	\end{algorithmic}
	\label{algo:alm2mapS2HATParallelCUDA} 
\end{algorithm}                 
                                
First, we partition the vector of the coefficients \aalm{} in global
memory into small tiles. In this way a block of threads can fetch
resulting segments in a sequence and fit them in shared memory. Next,
we calculate \blm{} values in the same way as in the \map2alm
algorithm. With all the necessary data in fast shared memory, in the
following steps, each thread evaluates in parallel an associated
Legendre function and updates the partial result $\bm{\Delta}^{A}_m$.

There are two major differences with the \map2alm algorithm discussed
earlier. First, the updates are performed without the race condition
problem as each thread computes, stores, and transfers to global
memory values which are specific to it and independent on the values
calculated by the other threads. Second, there is only one value per
thread, and therefore at most ${\tt NTB}$ values per thread block,
which need to be sent from any SM to global memory, what is typically
much smaller than the ${\cal O}(\ell_{max})$ volume required in the
case of \map2alm. We will return to these conclusions while discussing
our numerical experiments later on in this paper.

\subsubsection{Specific optimisations for GPU.}
In addition to the optimisations specific to our application and
described in the previous section, the algorithms developed here were
optimised following the general programming rules and guidelines as
developed for CUDA GPGPUs, e.g.,~\cite{Farber:2011,Sanders:2010}. For
completeness we briefly list the most consequential of them here. We
have limited the communication between the host and the device to the
minimum, with the exception of our alternative solution to the race
condition problem for \map2alm. We have expressly avoided asymmetrical
branching in the control flow on the GPU itself. We have minimised the
need to access the global memory by introducing buffering and
recalculating or reusing data, e.g., see
Algorithm~\ref{algo:alm2mapS2HATParallelCUDA}.  We have also removed,
whenever possible, branches in performance-critical sections to avoid
warp serialisation.

We note that this kind of optimisations have been studied in the case
of the inverse SHT performed on a single GPU by~\cite{hupca2012}. Our
experiments essentially confirm their findings, which we found
applicable to both direct and inverse transforms.  We therefore refer
the reader to their work for a detailed discussion.
 
\subsection{Exploitation of hybrid architectures}

\begin{figure}[htp]
  \begin{center}
    \includegraphics[width=0.75\textwidth]{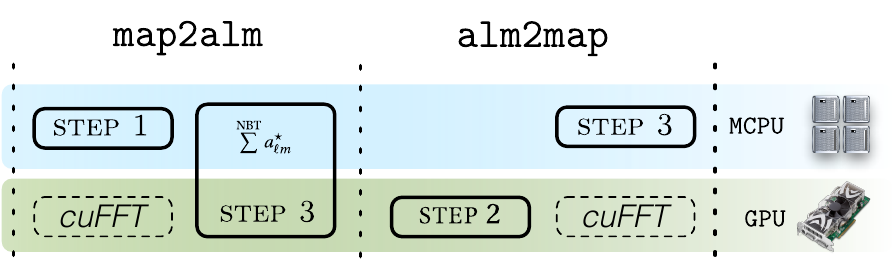}
  \end{center}
  \caption{Default assignment of the major steps of the SHT algorithm
    to one of the two architectures of a heterogeneous system composed
    of multi-core processors and GPUs.  {\sc step} 3 in {\tt map2alm}
    algorithm may involve any of the two architectures due to the
    optional use of CPU for reducing partial results between the
    execution of CUDA kernels.}
  \label{fig:heterogeneus}
    \vskip -0.5truecm
\end{figure}

Many new supercomputers are heterogeneous, combining computational
nodes interconnected with multiple GPUs.  This way, one or more
multi-core processors have access to one GPU. In such a case, the
front end user of the spherical harmonic package will have to decide
which type of architecture to choose for the calculation. The
three-stage structure of our algorithms allows mapping different steps
of the algorithms on either multi-core processors or accelerators.
The intention of this flexibility is to be able to assign each step of
the algorithm to the architecture on which its performance is better.
For example, Nvidia provides its own library for computing FFTs on
GPUs i.e., {\tt CUFFT} \cite{CUFFT}, whose efficiency strongly depends
on the GPU device and on the version of the routine.  However, there
are cases when highly tuned FFTs perform better if they are executed
on multi-core processors.  Note that our algorithms require the
evaluation of FFTs on vectors of complex numbers, whose length may
vary from ring to ring.  The default assignment for both transforms is
depicted in Figure~\ref{fig:heterogeneus}.  Due to their sequential
nature, we omit steps in which we evaluate the starting values of the
recurrence described in equation~\eqref{eqn:pmm} ({\sc step} 2 in
Algorithm \ref{algo:map2almBasic} and {\sc step} 1 in Algorithm
\ref{algo:alm2mapBasic}), which are always computed on the CPU.


\section{Numerical experiments}\label{sec:experiments} 

In order to validate our new algorithms as well as to evaluate their
performance in terms of their accuracy, scalability, and efficiency, we
conduct a number of experiments involving a comprehensive set of test
cases with a different, assumed geometry of the 2-sphere grid and
different values of $\ell_{max}$ and $m_{max}$.  Each experiment
involves first an inverse and then a direct SHT, and uses as input a
set of uniformly distributed, random coefficients,
$\bm{a}^{init}_{\ell m}$, with values in the range $(-1,1)$.  To
measure the accuracy of the algorithm we calculate the error as a
difference between the final output $\bm{a}^{out}_{\ell m}$ and the
initial set $\bm{a}^{init}_{\ell m}$ via the following formula,
\begin{eqnarray}\label{eq:error}
	{\cal D}_{err} & = & \sqrt{\frac{\sum\limits_{\ell}\sum\limits_{m}\l|\bm{a}^{init}_{\ell m} - \bm{a}^{out}_{\ell m}\r|^{2}}{\sum\limits_{\ell}\sum\limits_{m}\l|\bm{a}^{init}_{\ell m}\r|^{2}}}.
\end{eqnarray}
In these experiments we use HEALPix grids with a different resolution
parameter $N_{side}$, defining the number of the grid points/pixels,
$n_{pix} = 12N^{2}_{side}$, the number of the rings ${\cal
  R}_{N}=4N_{side}-1$, and the maximum number of points along the
equatorial ring, $4N_{side}$. Finally, we also set
$\ell_{max}=m_{max}=2 \, N_{side}$, except for some specific
experiments.  In the following we compare our algorithms with the
transforms as implemented in a publicly available library called
\textbf{libpsht}~\cite{libpsht}. To the best of our knowledge it is
the fastest implementation of SHT at this time suitable for
astrophysical applications, and it is essentially a realisation of the
same algorithm as the one used in our parallel algorithms. The
\textbf{libpsht} transforms are highly optimised using explicit
multithreading based on OpenMP directives, and SIMD extensions (SSE
and SSE2), but its functionality is limited to shared memory systems.
In all our tests we set the number of OpenMP threads, for both
\textbf{libpsht} and our multi-threaded code, to be equal to the
number of physical cores, i.e., $\text{OMP\_NUM\_THREADS} = 4$.

\subsection{Scaling with test case sizes}
\label{sssub:validate}

In this set of experiments we generate a set of initial coefficients,
$\bm{a}^{init}_{\ell m}$, for different values of $\ell_{max}$ and
$N_{side}$, and use them as an input for our two implementations of
the transforms (MPI/OpenMP and MPI/CUDA) as well as for those included
in the \textbf{libpsht} library.  Our experiments are performed on
three different computational systems listed in Table~\ref{tab:arch}
and are run either on one multi-core processor or on a CPU/GPU node.

\begin{table}[h]
	\begin{center}
		\begin{tabular}{lcll} 
		  \emph{CPU}  & \emph{clock speed} & \emph{GPU}  & \emph{Compilers}  \\ \hline\hline
			{\sc Core i7-960K}  &  3.20 GHz & {\sc GeForce GTX 480 } &  gcc 4.4.3 / nvcc 4.0   \\
			{\sc Core i7-2600K} &  3.40 GHz & {\sc GeForce GTX 460 } &  gcc 4.4.5 / nvcc 4.0   \\
			{\sc Xeon X5570}    &  2.93 GHz & {\sc Tesla S1070 }     &  icc 12.1.0 / nvcc 4.0     \\ \hline
		\end{tabular}
	\end{center}
\caption{Different Intel multi-core processors and NVDIA GPUs used in our experiments.}\label{tab:arch}
\end{table}
 

\begin{figure}[htp]
  \begin{center}
    \includegraphics[width=0.95\textwidth]{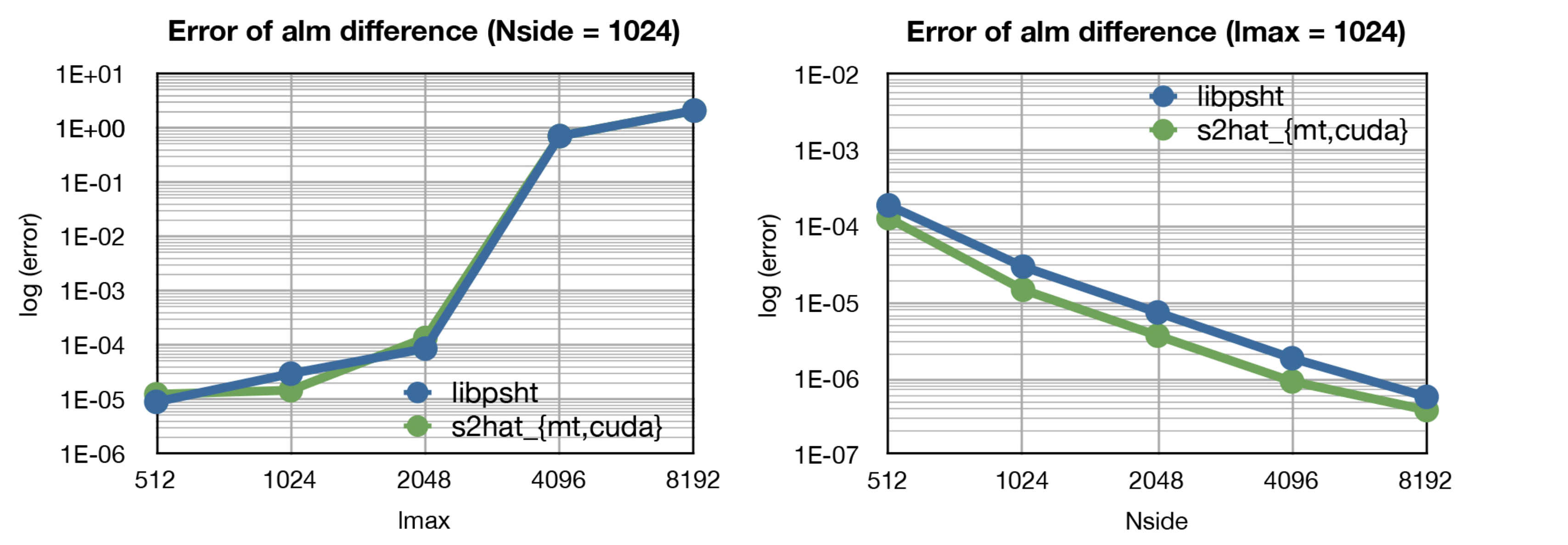} 
  \end{center}
  \caption{Relative error, ${\cal D}_{err}$, for
    \textbf{s2hat\_\{mt,cuda\}} and \textbf{libpsht} transform pairs
    on the HEALPix grids for different values of $\ell_{max}$ and
    $N_{side}$. Data points show the error values found in our
    tests. The loss of precision seen in the left panel for
    $\ell_{max} \simgt 2048$ and shared by both implementations is due
    to aliasing of the high-$\ell$ power contained in the spatial
    modes, which are not longer supported by the grid. }
  \label{fig:libpsht_err}
\end{figure}

Figure~\ref{fig:libpsht_err} displays the error ${\cal D}_{err}$ for
\textbf{s2hat\_\{mt,cuda\}} and \textbf{libpsht} transform pairs
computed for a set of $\ell_{max}$ and $N_{side}$ values on the
HEALPix grid. Every data point corresponds to the error value
encountered in the pairs of transforms. It can be seen that all the
implementations provide almost identical precision of the solution in
all the cases considered here.

\begin{figure}[htp]
  \begin{center}
    \includegraphics[width=0.95\textwidth]{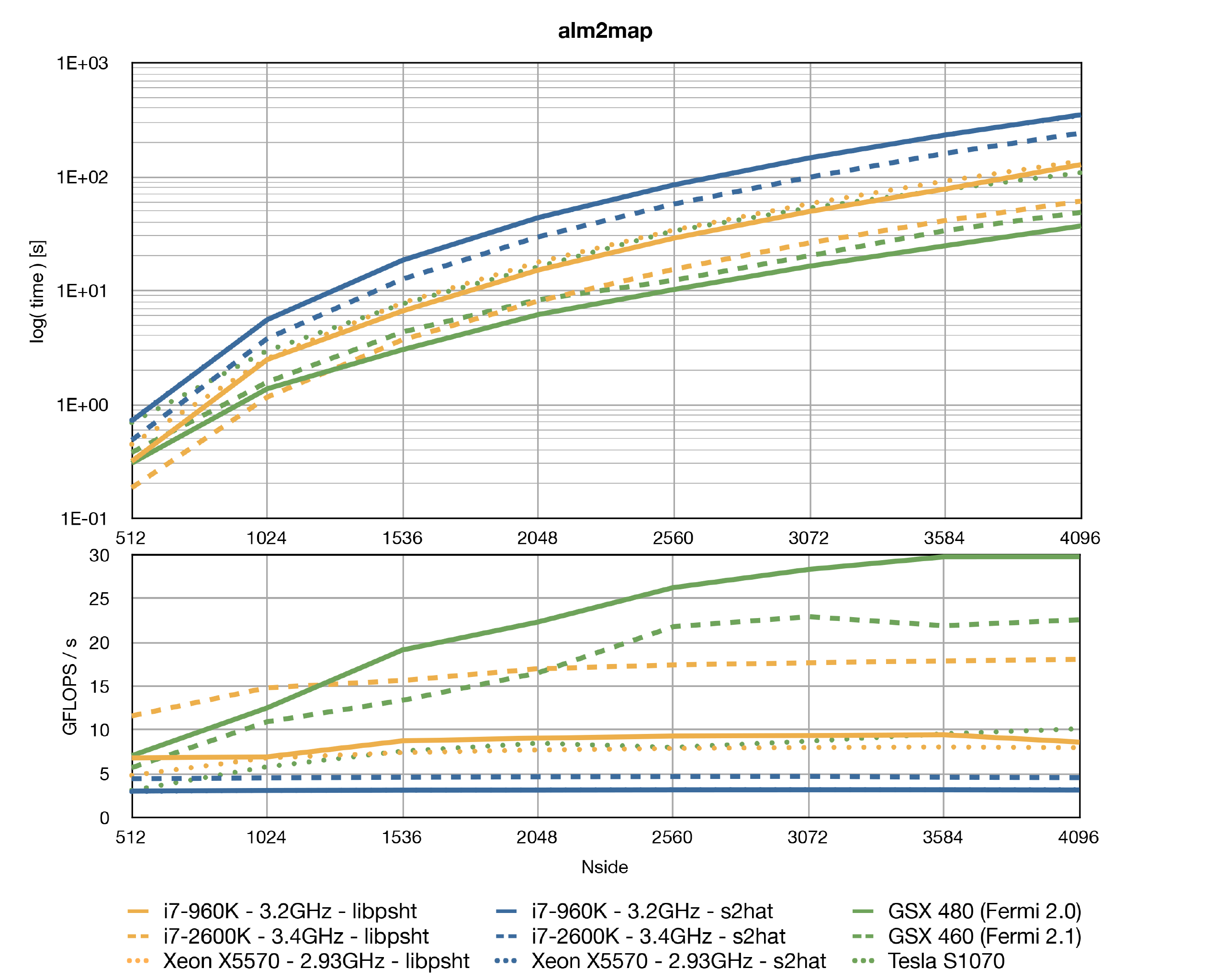}
  \end{center}
  \caption{Benchmarks for the different implementations of the inverse
    SHT performed on either Intel multi-core processors or GPUs, see
    Table~\ref{tab:arch} for details.  Results of the \textbf{libpsht}
    code are compared with those of the hybrid \s2hat codes (orange,
    blue and green lines, respectively) and are obtained for the
    HEALPix grid with $N_{side}=\ell_{max}/2$. The top plot shows
    timings for calculating $\bm{\Delta}_m^A$,
    equation~\eqref{eqn:DeltaDef}, involving the recurrence to
    calculate the associated Legendre functions. In the case of the
    CUDA routines, the numbers include time of data movements between
    CPU and GPU. The bottom plot shows the number of GFlops per second
    i.e., the total number of flops required in the algorithms divided
    by the CPU time (see table~\ref{tab:flops}). Note that the CUDA
    routines loose performance in comparison to the SSE optimised
    \textbf{libpsht} due to very costly reduction of partial results,
    section~\ref{sub:sht_on_gpu}.  Nevertheless, Nvidia GTX 480, green
    solid line, outperforms \textbf{libpsht} on the multicore
    processor Core i7-960K, solid yellow line. The dotted blue line
    overlaps perfectly with the solid blue line in both
    plots.}\label{fig:libpsht_ndside_alm2map_t}
\end{figure}

\begin{figure}[htp]
  \begin{center}
    \includegraphics[width=0.95\textwidth]{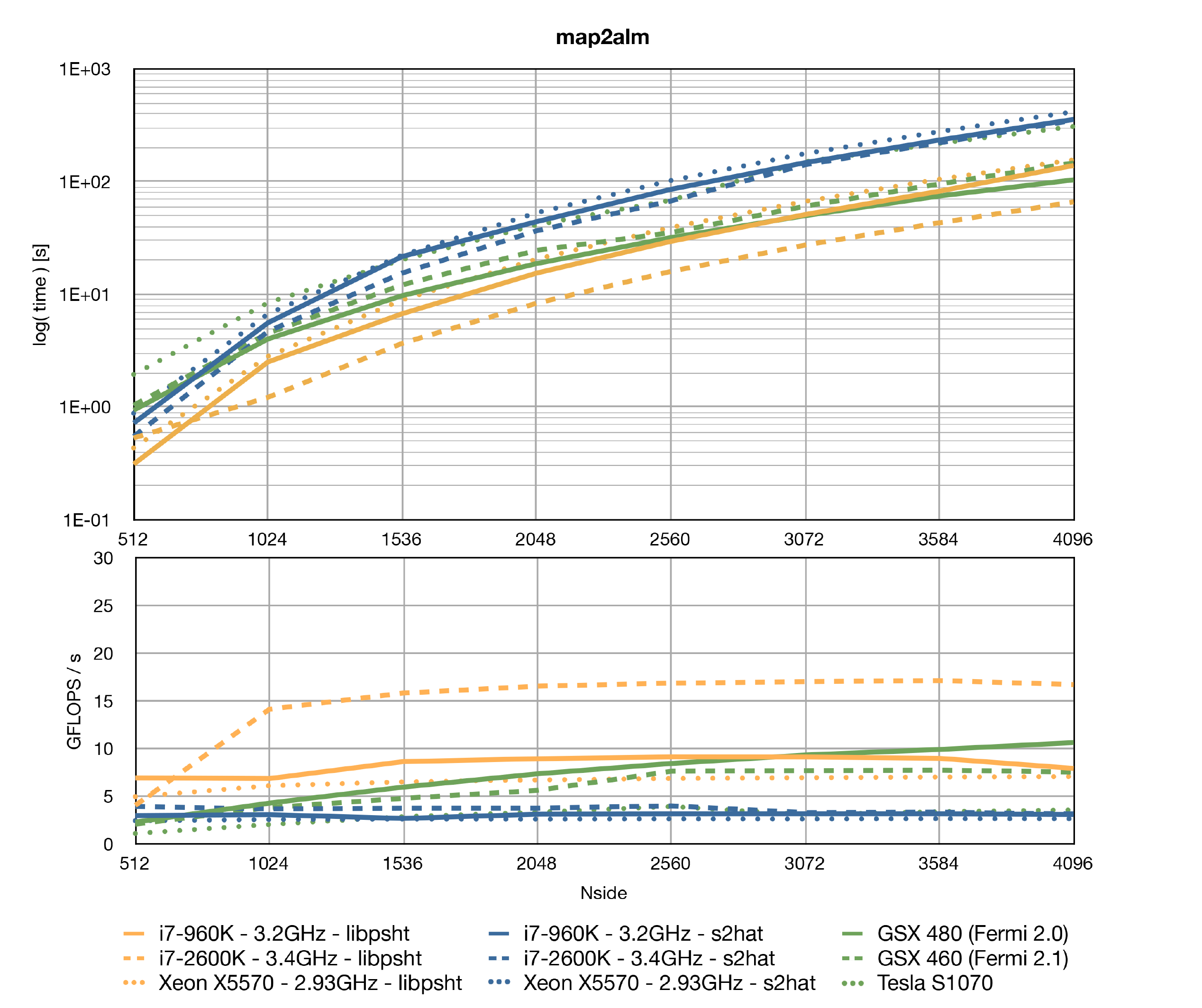}
  \end{center}
  \caption{Performance of the implementations of the direct SHT on the
    systems described in Table~\ref{tab:arch}. The top plot shows the
    time needed for the calculation of $\bm{\Delta}^S_m$ including the
    ${\cal P}_{\ell m}$-recurrence, equation~\eqref{eqn:almfftdef}.
    The bottom plot displays the achieved numbers of GFlops per
    second.  For all the runs we use the HEALPix grid with
    $N_{side}=\ell_{max}/2$.  Orange corresponds to the
    \textbf{libpsht} routine, while blue and green to the hybrid
    \s2hat ones, multi-threaded and CUDA, respectively.
  }\label{fig:libpsht_ndside_map2alm_t}
\end{figure}

Figures~\ref{fig:libpsht_ndside_alm2map_t}
and~\ref{fig:libpsht_ndside_map2alm_t} display the performance of the
recurrence part in our algorithms ({\sc step} 3 in Algorithm
\ref{algo:map2almBasic} and {\sc step} 2 in Algorithm
\ref{algo:alm2mapBasic}) in terms of measured time in seconds and
number of GFlops per second. The number of flops includes all the
operations necessary for maintaining the numerical stability of the
recurrences, e.g., the rescaling in the case of \s2hat routines as
described in section~\ref{sec:background}.  The resolution of the
problem progressively increases from left to right, as do the data
objects, memory load, and communication volume,
section~\ref{sec:perfan}.  The maximal value of $N_{side} = 4096$ was
set due to the limitation of the memory of the single GPU used in
these tests.  We note that for the CUDA routines, the measurements
include the time spent in data movements between GPU and CPU.

The expected number of flops in these tests is proportional to
$N_{side}^3$, and such a scaling is indeed closely followed by 
the experimental results obtained for the multi-threaded \s2hat transforms, which are depicted by
the blue lines in the top plot of Figures~\ref{fig:libpsht_ndside_alm2map_t}
and~\ref{fig:libpsht_ndside_map2alm_t}. The other curves in these plots are
however somewhat flatter. This is because these routines become more efficient for the
larger data sets.
This fact is further emphasised in the bottom plots of these two figures, where the
performance of the transform routines is shown in terms of
GFlops per second.  The lowest, blue curves, corresponding to multithreaded \s2hat 
transforms remain nearly constant across the
entire tested range of $N_{side}$, while the other implementations, and in particular
the \s2hat-GPU ones, tend to improve their overall efficiency with the
size of the test.

The fact that multithreaded \s2hat transforms are systematically
slower than \textbf{libpsht} by a constant factor could have been
expected  due to the lack of SSE/AVX intrinsics in our
code. These allow performing two or more arithmetic operations of the
same kind with a single machine instruction on some hardware devices.
Moreover, on the latest intel i7-2600K processor, SSE-optimised
\textbf{libpsht} implementation further benefits from the new Intel AVX
128-bit instructions set, which have lower instruction latencies, than 
 the standard SSE extension. We note that additional
possible optimisation of both libraries could take advantage of
extended width of the SIMD register and Advanced Vector Extensions
(AVX), which enable operating on $8$ float elements per iteration
instead of a single one.  For the GPU transforms, the following
observations can be made:
\begin{itemize}
\item The direct transform on CUDA is slower, not only with respect to
  the corresponding CPU version, but also in comparison with the
  inverse transform on the same architecture. This is in spite of the
  fact that on the algebraic level, the calculations, and their
  implementations, of the objects $\bm{\Delta}_m^{A/S}$ are
  essentially the same for both directions of the transform.  The
  performance loss is due to the reduction of partial results, and is
  incurred first in the shared memory, then in either the slow global memory
  or  the host memory of the CPU, as already discussed on the
  theoretical level in section~\ref{sub:sht_on_gpu}.
\item 
Our code on the two Fermi GPUs (GTX 460 \& GTX 480) outperforms
\textbf{libpsht} run on the latest Intel i7 processor in the
evaluation of the inverse transform.
\item The CUDA routines seem to favour cases with large $N_{side}$
  which allow the usage of many threads (as we explained in section
  \ref{sub:sht_on_gpu}, each ring of the sphere is assigned to one
  CUDA thread) i.e., both GPU routines gain performance with growing
  number of sky-pixels (thus, also the rings).
\end{itemize}

Recapitulating, for the inverse transforms we find that the CUDA
routine performs better than some of the popular routines for
multi-core processors, even if SSE optimisations are used to
enhance the performance in these latter cases. For CPUs, the direct and
inverse transforms are shown to display similar performance. This is
however not the case for the GPUs, where the direct transform is found
systematically slower than the inverse one. Nonetheless, the direct
transform on GPUs still runs comparably fast to that run on at least
some of the current multi-core processors. In both these cases the
performance of the CUDA transforms with respect to the multi-core
ones, keeps on improving with the sizes of the studied problems.

\subsection{Scaling with number of multi-core processors and GPUs}

In this section we discuss the performance of our algorithms on
clusters of hybrid computers.  As before, we set $\ell_{max}= 2\,
N_{side}$ in all the tests.  The results are obtained on CCRT's
supercomputer Titane, one of the first hybrid clusters built, based on
Intel and Nvidia processors. It has $1,068$ nodes dedicated to
computation, each node is composed of $2$ Intel Nehalem quad-core
processors ($2.93$ GHz) and $24$ GB of memory.  A part of the system
has a multi-GPU support with $48$ Nvidia Tesla S1070 servers, $4$
Tesla cards with $4$ GB of memory and $960$ processing units are
available for each server. Two computation nodes are connected to one
Tesla server by a PCI-Express bus, therefore $4$ processors handle $4$
Tesla cards. This system provides a peak performance of $100$ Tflops
for the Intel part and $200$ Tflops for the GPU part. The Intel
compiler version 12.0.3 and the Nvidia CUDA compiler 4.0 are used for
compilation. Furthermore, we define the MPI processes in our
experiment such that each of them corresponds to a pair of one
multi-core processor and one GPU (4 MPI processes per one Tesla
server). To increase intra-node parallelism, we set-up for all runs
the number of OpenMP threads equal to the number of physical cores of
the Intel Nehalem processor ($\text{OMP\_NUM\_THREADS} = 4$).

\begin{figure}[htp]
  \begin{center}
    \includegraphics[width=0.95\textwidth]{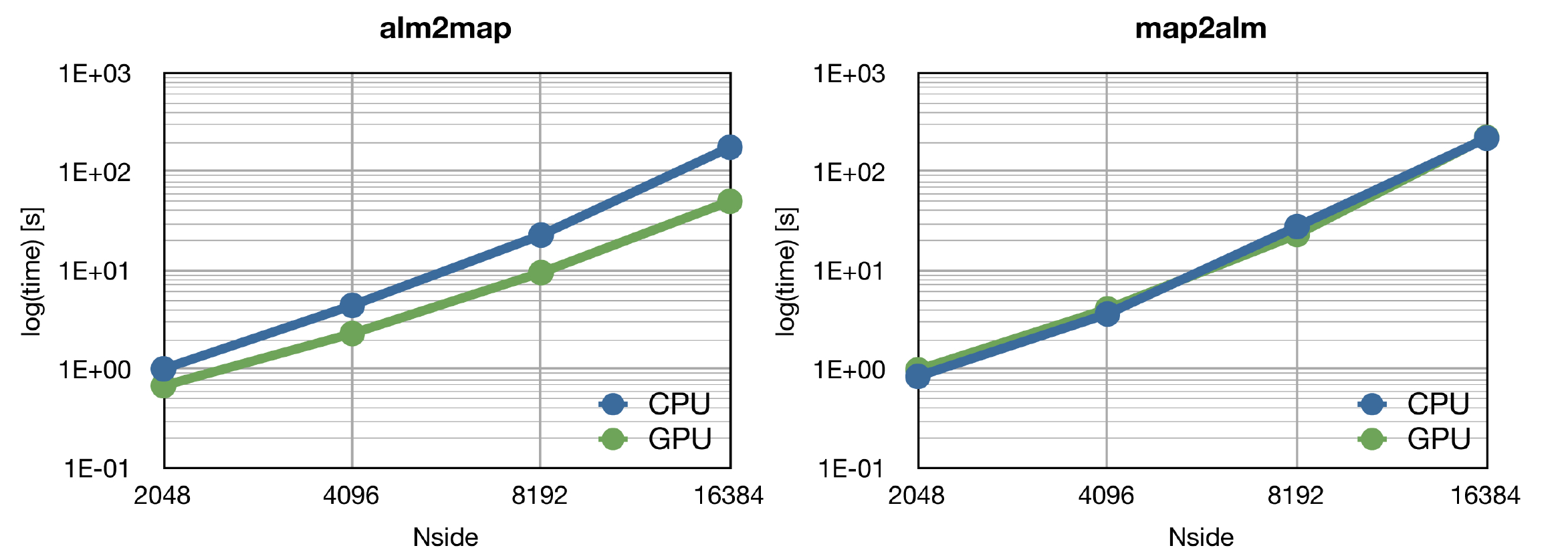}
  \end{center}
  \caption{Runtime of the inverse (left), and direct (right),
    transforms as a function of $N_{side}$, ($\ell_{max} = 2\,
    N_{side}$). The experiments use $128$ MPI processes corresponding
    to $128$ pairs of Intel Nehalem quad-core processors and Nvidia
    Tesla S1070.}\label{fig:scale_all_nside}
\end{figure}

Figure~\ref{fig:scale_all_nside} shows the performance on 128 hybrid
processors of the computation of a pair of the SHT as a function of
the problem size.  It can be seen that our both implementations scale
well with respect to the problem size, and in all the cases follow
closely the predicted scaling $\propto N_{side}$.  As already noted
in the previous section, only for the inverse transforms we see a
performance gain when using GPUs.

\begin{figure}[htp]
  \begin{center}
    \includegraphics[width=0.95\textwidth]{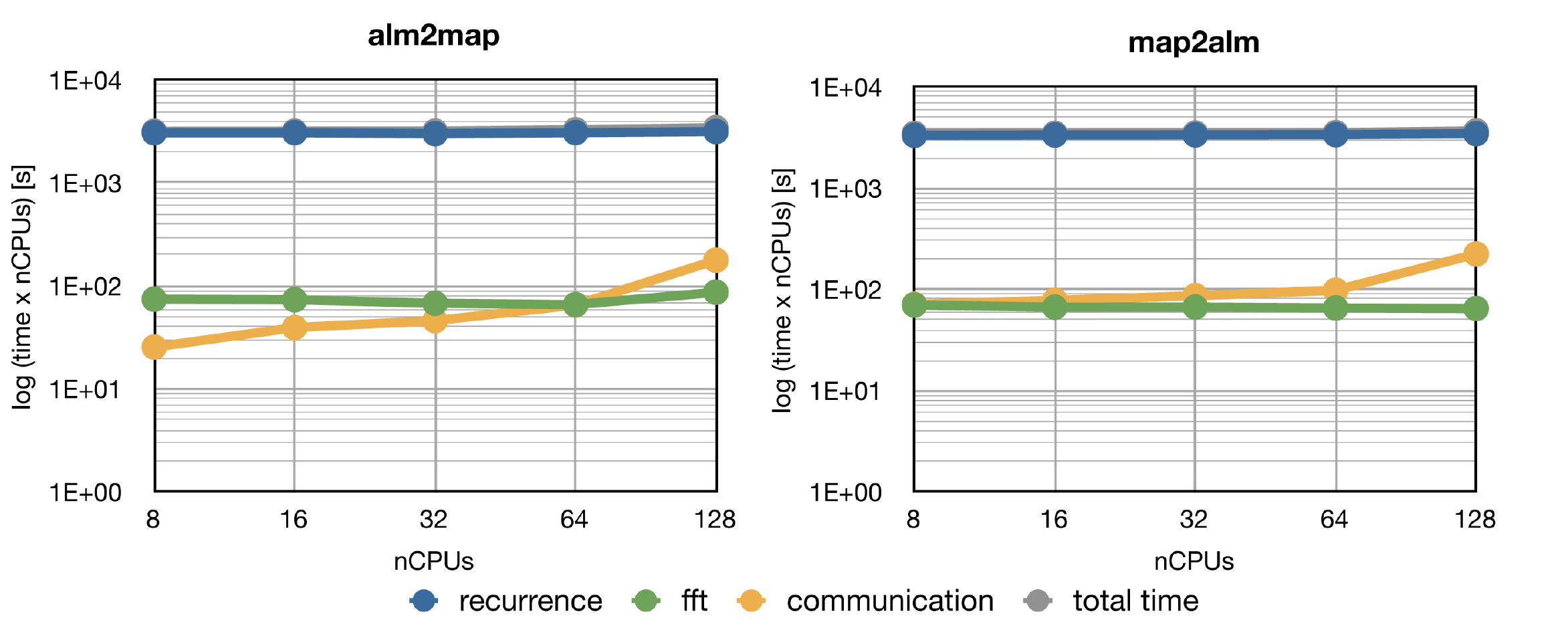}
  \end{center}
  \caption{Time breakdown into the main operations for the two
    transforms, \textbf{alm2map\_mt}, left, and \textbf{map2alm\_mt},
    right, in the hybrid MPI/OpenMP case. The results are obtained for
    experiments with $N_{side} = 16384$ and $\ell_{max} =
    2\,N_{side}$.}\label{fig:scale_all_cpu}
\end{figure}

\begin{figure}[htp]
  \begin{center}
    \includegraphics[width=0.95\textwidth]{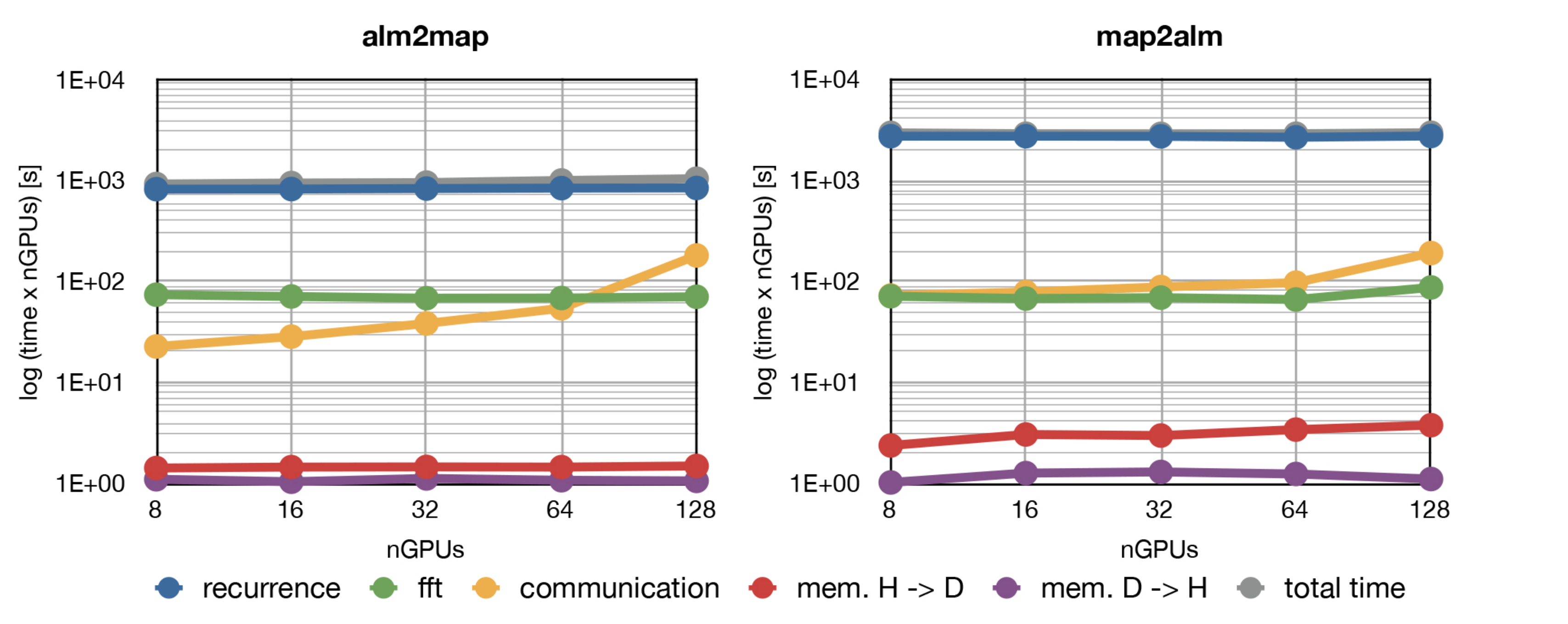}
  \end{center}
  \caption{As Figure~\ref{fig:scale_all_cpu}, but for the hybrid
    MPI/CUDA case.} \label{fig:scale_all_gpu}
\end{figure}

\begin{figure}[htp]
  \begin{center}
    \includegraphics[width=0.95\textwidth]{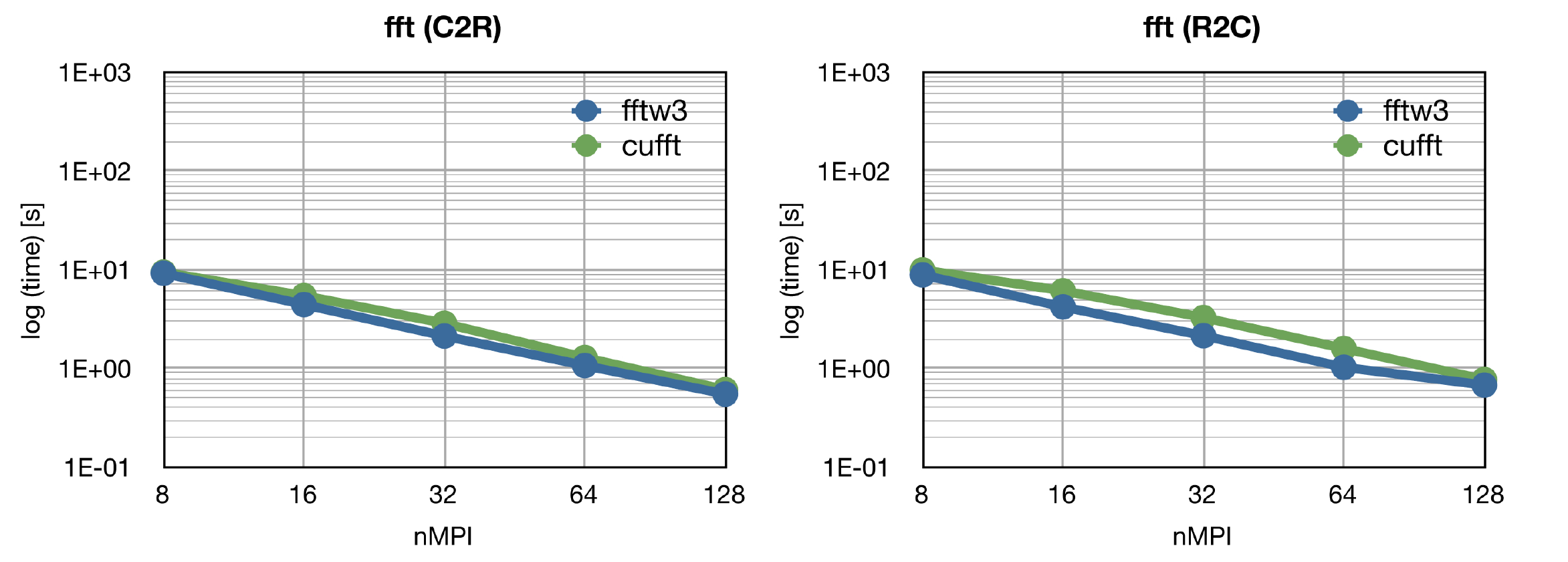}
  \end{center}
  \caption{Comparison of two FFT libraries on the Titane
    supercomputer: CUFFT on GPU and Intel MKL Fourier Transform on CPU. The results correspond to
    the experiment with $N_{side} = 16384$ and $\ell_{max} = 2\,
    N_{side}$.}\label{fig:cufft_vrs_fftw3}
\end{figure}

Figures~\ref{fig:scale_all_cpu} and~\ref{fig:scale_all_gpu} depict the
runtime breakdown into the main steps of the algorithm for both
directions of the transforms in the case of a fixed size of the
problem. The shown times are multiplied by the number of MPI
processes.  For both transforms, all major operations scale nearly
perfectly in the regime studied here. Indeed, as expected and
discussed in section~\ref{sec:perfan}, the communication cost grows
slowly with the number of processes.  As predicted earlier on,
section~\ref{sub:t_consumption}, the evaluation of the objects,
$\bm{\Delta}_m^{A/S}$, dominates by far over the remaining steps
determining the overall run time. In the contrary, the time spent on
memory transfers between the host and the GPU devices never exceeds
$1$ second.  In both implementations we assign the evaluation of FFTs
to the CPU due to its slightly better performance in comparison to
CUFFT also available on Titane (see Figure~\ref{fig:cufft_vrs_fftw3}).

\begin{figure}[htp]
  \begin{center}
    \includegraphics[width=0.95\textwidth]{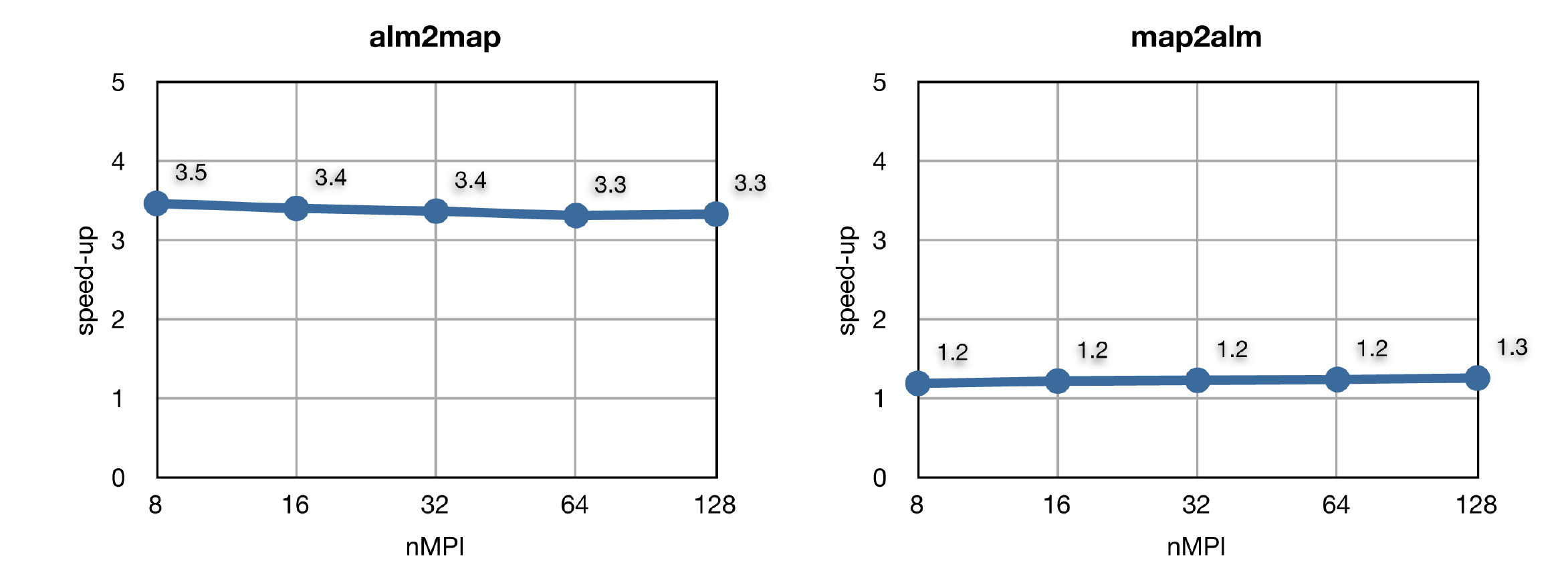}
  \end{center}
  \caption{Average speed-up of the hybrid implementation MPI/CUDA with
    respect to the hybrid MPI/OpenMP version for different number of
    MPI processes.}\label{fig:scale_sup}
\end{figure}

Finally, in terms of overall runtime, the implementation of
\textbf{alm2map} for CUDA is, for all our big test cases, on average
$\times 3.3$ faster than the multithreaded version. In the contrary,
due to the time consuming reduction operations, \textbf{map2alm} is
only slightly faster than its CPU-based counterpart. An average
speed-up of the hybrid implementation MPI/CUDA with respect to the
hybrid MPI/OpenMP version is depicted in Figure~\ref{fig:scale_sup}.
We emphasise that no SSE intrinsics have been used in the
multi-threaded routines used in these tests. In addition, the Nvidia
GTX devices used here had double-precision throughput purposefully
degraded by the manufacturer to 25\% of the full design performance
(when compared to Tesla card family with the same chipset).


\begin{figure}
  \centering
  \subfigure[Reference map evaluated on CPU]{\label{fig:ref}\includegraphics[width=0.45\textwidth]{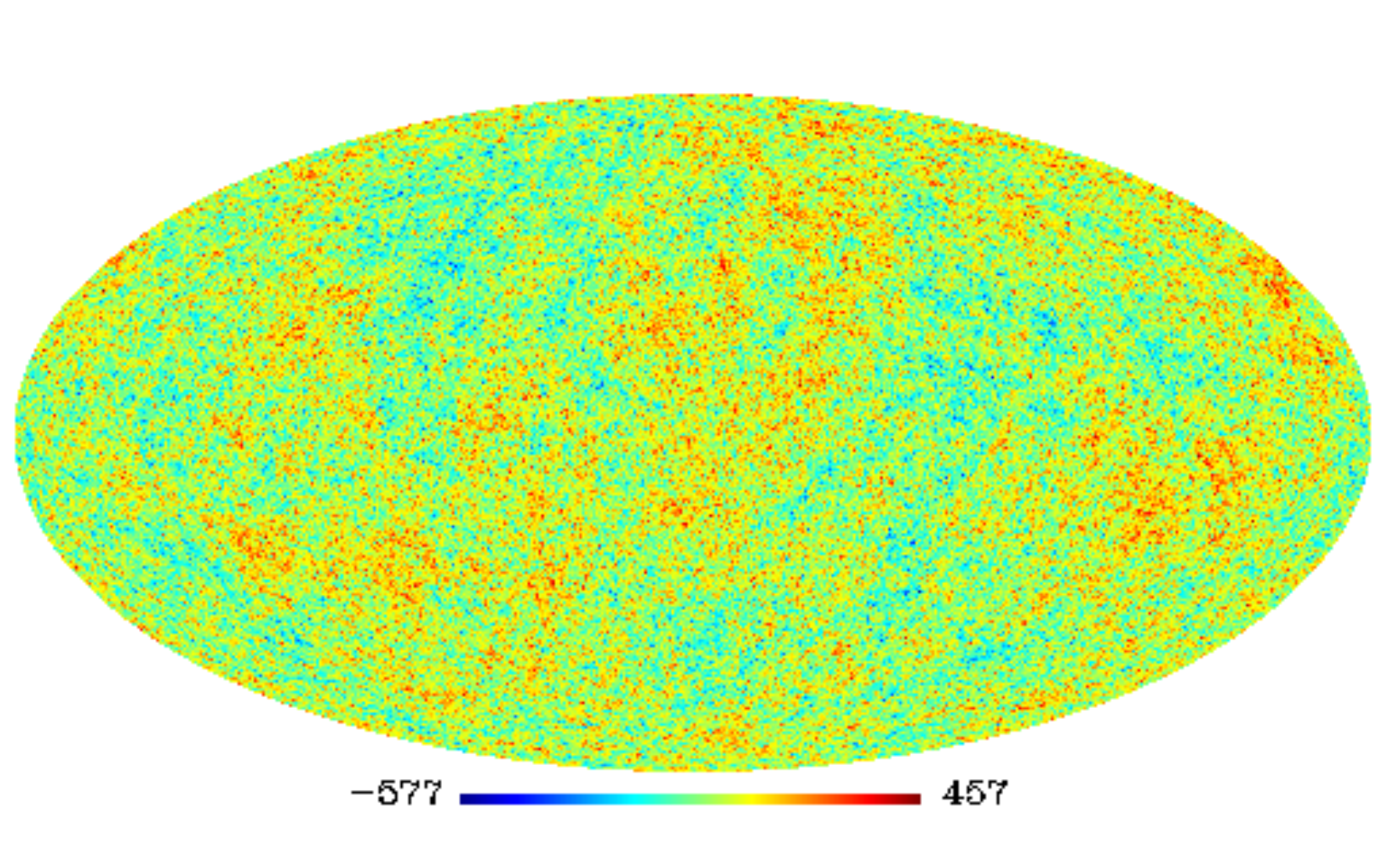}}                
  \subfigure[Map generated on Nvidia Tesla S1070 GPU]{\label{fig:cuda}\includegraphics[width=0.45\textwidth]{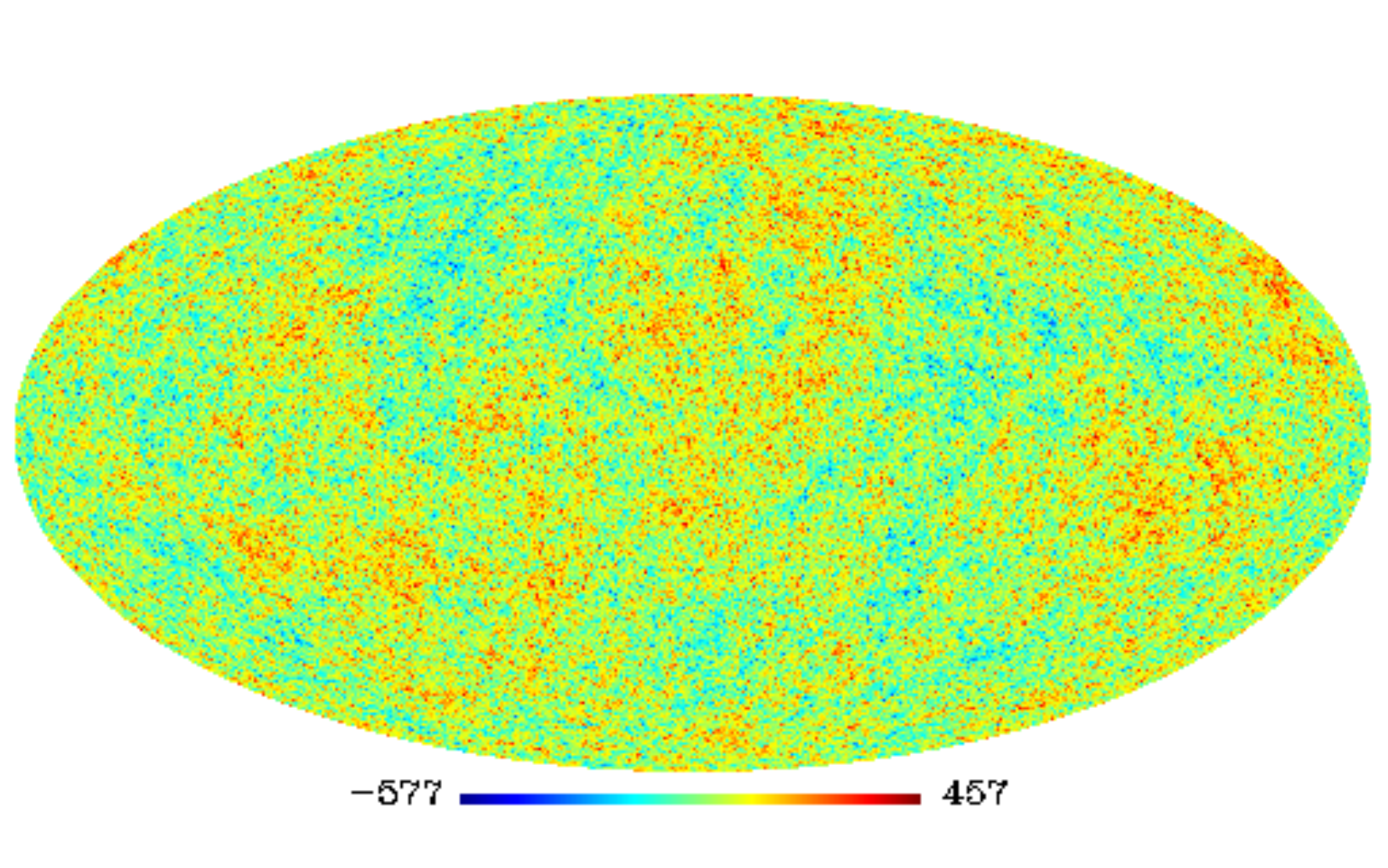}} \\
  \subfigure[Absolute difference between maps generated on CPU and Nvidia GPU.]{\label{fig:diff}\includegraphics[width=0.45\textwidth]{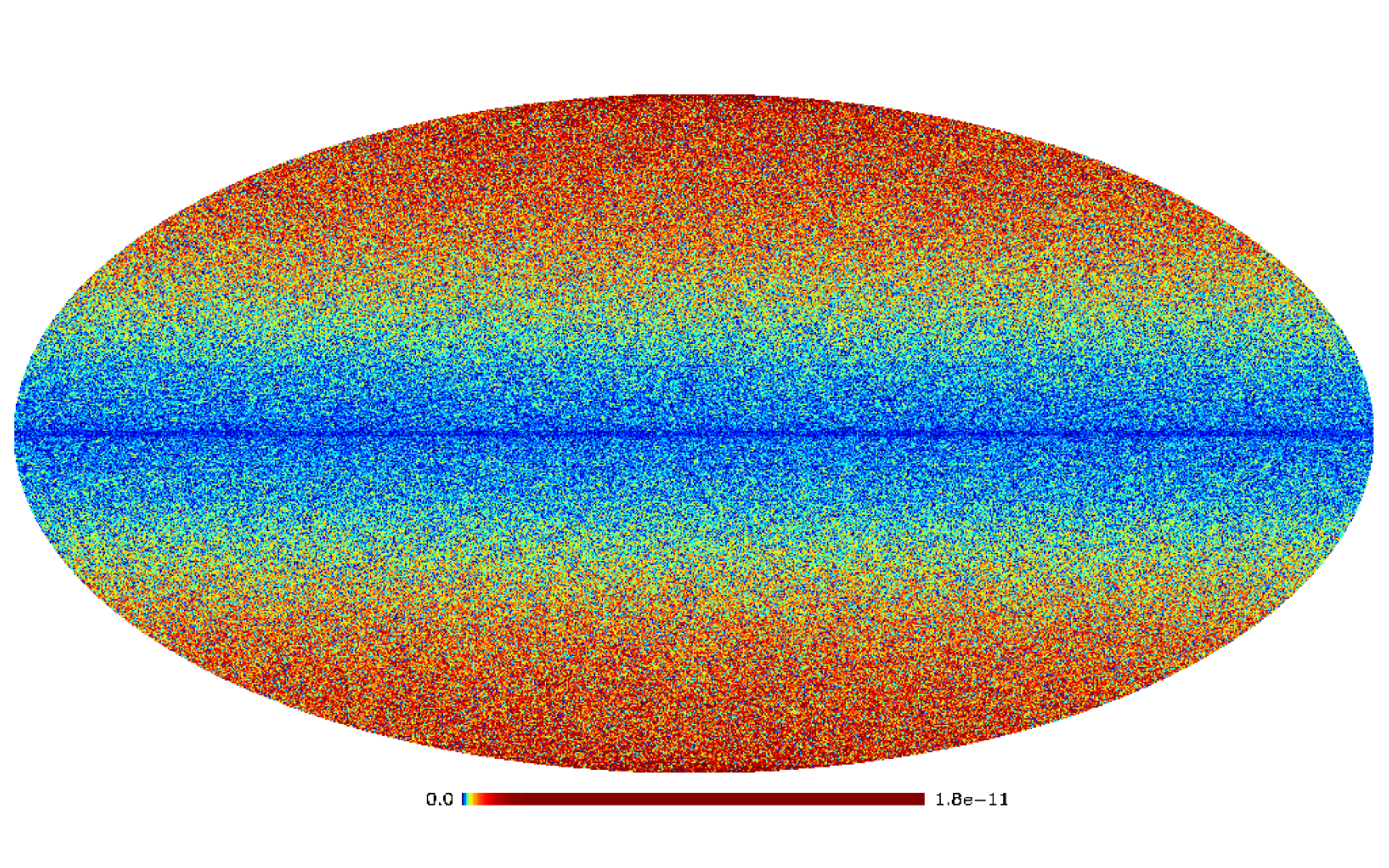}}
  \caption{CMB maps generated on two different architectures and their absolute difference.}
  \label{fig:cuda_cpu_diff}
\end{figure}

To assess the accuracy of results on the different architectures used
in this paper, we perform a serie of experiments in which we convert
the same random set $\bm{a}^{init}_{\ell m}$ (see beginning of this
section) to a HEALPIx map on two different architecture, namely Nvidia
GPU and x86 CPU.  Figure~\ref{fig:cuda_cpu_diff} displays the absolute
difference of those two maps.  In all our tests the maximum absolute difference
between two values of the sky signal recovered in the same pixel never exceeded $\sim 10^{-11}$,
indicating a very good agreement between these two architectures.

\section{Conclusion and future work}\label{sec:conclusions} 
This paper describes parallel algorithms for computing spherical
harmonic transforms employing two variants of intra-node parallelism,
specific to two different architectures i.e., multi-core processors
and GPUs, and an optional, distributed memory layer based on MPI
directives. We present a detailed discussion of the developed
algorithms and perform comprehensive sets of tests comparing them
against each other, as well as against an independent implementation
of the transforms included in the \textbf{libpsht} package.  We then
evaluate their overall efficiency and scalability.  In particular, we
show that our inverse SHT run on GeForce 400 Series GPU equipped with
latest CUDA architecture outperforms {\bf libpsht}, a state of the art
implementation for multi-core processors which benefits from the usage
of SSE intrinsics, and executed on the latest Intel Core i7-2600K
processor. At the same time the CUDA direct transform, though slower
than its inverse-direction counterpart, provides a comparable, though
typical lower, performance to that of {\bf libpsht}.  For the runs
including the MPI layer, we find that a cluster of Nvidia Tesla S1070
can accelerate an overall execution of the inverse transform by as
much as $3$ times with respect to the hybrid MPI/OpenMP version run on
the same number of quad-core processors Intel Nehalem (no SSE
intrinsics). The gain for the direct transforms is again found however
to be significantly more modest and only about $20-30$\%.  We discuss
in details the specific features of the nontrivial algorithms invoked
for the spherical transforms calculations and elucidate the sources of
the inverse -- direct transforms dichotomy, which prevent a full
adaptation of the latter algorithm to the CUDA architecture.

The current implementation leaves some space for future
optimisations. In particular, as a future work, we could add SSE
intrinsics ( or/and AVX for the latest x86
architectures) to our routines in order to capitalise on a low level
vectorization on CPUs. On the algorithmic side and from the
application perspective, it would be important to add 
support of transforms of objects with non-zero spin, which are
commonly encountered in the CMB science.


\section{Acknowledgments}

This work has been supported in part by French National Research Agency (ANR) through COSINUS program (project MIDAS no. ANR-09-COSI-009).
The HPC resources were provided by GENCI- [CCRT] (grants 2011-066647 and 2012-066647) in France and 
by the NERSC in the US, which is supported
by the Office of Science of the U.S. Department of Energy under Contract No. DE-AC02-05CH11231. Some of the results in this paper have been derived using the HEALPix package \cite{Gorski_etal_2005}.


\bibliographystyle{wileyj}  
\bibliography{s2hat_cpe_format_1.1.1}

\end{document}